\documentclass{article}
\usepackage[T1]{fontenc}
\usepackage[utf8]{inputenc}
\usepackage{microtype}
\usepackage{graphicx}
\usepackage{booktabs,array,amsmath,amssymb}
\usepackage{xcolor}
\usepackage{courier,fix-cm}
\usepackage{url}
\usepackage[preprint,nohyperref]{mlsys2025}
\usepackage{placeins}
\usepackage{float}
\usepackage{dblfloatfix}
\setcitestyle{citesep={;}}
\ifdefined
\else

\fi
\mlsystitlerunning{Autoformalization for Ambiguous Inputs}
\begin{document}
\twocolumn[
\mlsystitle{Autoformalization for Ambiguous Inputs\\Precision, Coverage, and Cost across SQL and Logical Forms}
\begin{mlsysauthorlist}
\mlsysauthor{Henry Shen}{yh,stevens}
\mlsysauthor{Xinling Liao}{yh,kcl}
\mlsysauthor{Adarsh Vatsa}{stevens}
\mlsysauthor{William Eiers}{stevens}
\end{mlsysauthorlist}
\mlsysaffiliation{yh}{YH Intelligence Technology, Co., Ltd.}
\mlsysaffiliation{stevens}{Stevens Institute of Technology}
\mlsysaffiliation{kcl}{King's College London}
\mlsyskeywords{Autoformalization, ambiguity, SQL, logical forms, language model workflows}
\vskip 0.3in
\begin{abstract}
Agentic systems translate natural-language requests into executable artifacts while managing ambiguity and inference cost. We evaluate seven autoformalization workflows on SQL and first-order logic benchmarks for ambiguous inputs. Across five SQL workflows, the raw model leads with F1 of 53.35\% on AMBROSIA and 60.85\% on AmbiQT. On AmP, DRIFT gains 15.98 F1 points over the raw model at 2.03 times its token cost, whereas LoC uses 18.65 times the raw model's tokens with lower F1. Within AutoCedar, structured interpretation lowers F1 by 6.44--6.95 points and raises AMBROSIA token use 2.2-fold. Paired reviewer comparisons show modest Kimi K3 gains of 1.93 points on AMBROSIA and 2.04 on AmbiQT. Clear--ambiguous AMBROSIA analysis shows greater relative stability for AutoCedar, with a smaller F1 drop than the raw model (11.46 vs.\ 16.68 points), alongside lower aggregate F1 (50.16\% vs.\ 53.35\%). Autoformalization alone does not guarantee robustness to ambiguity: the tested additions offer no uniformly effective remedy, making accuracy, ambiguity handling, and inference cost joint design requirements.
\end{abstract}
]
{\let\thefootnote\relax\footnotetext{\textsuperscript{1}YH Intelligence Technology, Co., Ltd.\quad
\textsuperscript{2}Stevens Institute of Technology.\quad
\textsuperscript{3}King's College London.}}
\raggedbottom

\section{Introduction}
Agentic systems combine language models with tools to turn natural-language requests into actions over software and data. Database agents inspect schemas, generate and execute queries, and revise them using feedback \citep{lei2025spider2}. In prior work, access-control workflows translate requirements into policies governing which users or services may act on protected resources \citep{vatsa2026autocedar}. These applications place interpretation at an operational boundary: a query determines which records are retrieved, while an authorization policy determines who may read or modify them. Preserving the intended scope is essential to useful data access and privacy protection \citep{cutler2024cedar}.

Ambiguity can alter that scope before execution. A broad interpretation may include unintended records or permissions, while a narrow one may omit required results or deny legitimate access. A SQL query can execute successfully while committing to one of several meanings. Consider the request for singers' ``names, countries, and ages'' in Figure~\ref{fig:running-example}. The schema offers both nationality and citizenship for ``countries,'' and AmbiQT records a reference query for each.

\begin{figure}[!t]
\centering
\fbox{\begin{minipage}{0.94\linewidth}\small\color{black}
\textbf{AmbiQT 000005.} ``What are the names, countries, and ages for every singer in descending order of age.''\par\smallskip
\textbf{Two reference readings.} Countries denotes nationality or citizenship. The reference queries differ in the selected column:\par\smallskip
{\ttfamily\footnotesize SELECT name, nationality, age\par FROM singer ORDER BY age DESC;}\par\smallskip
{\ttfamily\footnotesize SELECT name, citizenship, age\par FROM singer ORDER BY age DESC;}\par\smallskip
\begin{tabular}{@{}lrrr@{}}
\toprule
Returned set & Precision & Recall & F1\\
\midrule
Direct (raw model) & 66.67 & 100.00 & 80.00\\
Roundtrip & 100.00 & 100.00 & 100.00\\
\bottomrule
\end{tabular}
\end{minipage}}
\caption{Coverage and precision can separate on the same ambiguous input. Both workflows recover the two reference answer classes; Direct also returns an additional prediction after grouping. Scores are percentages. Supplementary Table~\ref{tab:sample-evidence-10} gives the saved outputs and intermediate representations.}
\label{fig:running-example}
\end{figure}

A workflow can recover both readings yet also return additional answers, or return an internally consistent query that misses both references. This distinction matters in batch translation, autonomous agents, and offline requirements review, where an ambiguous input can reach execution before a user clarifies it.

We study ambiguity-aware formalization as recovery of a set of supported readings: one for a clear input and several for an ambiguous one. This set is useful both for automated processing and as the candidate set presented to a later clarification step. Methods such as ARTEMIS and Monty use human judgments to resolve competing specifications \citep{mendoza2026artemis,liu2026faithful}. Our experiments examine the preceding problem of producing candidates from the input and its context, using model feedback and formal checks.

The central question is how generation, checking, repair, and retrieval change the returned set. We evaluate seven workflows on three benchmarks: AMBROSIA and AmbiQT for SQL, and AmP for first-order logical forms \citep{saparina2024ambrosia,bhaskar-etal-2023-benchmarking,stengeleskin2024amp}. We then use AutoCedar as a common pipeline to study two placements of additional model information: interpretation before generation and feedback during review. We evaluate precision, recall, and token counts across the different modalities.

Four findings connect workflow design to the formal target, the placement of feedback, and input ambiguity:
\begin{enumerate}
\renewcommand{\labelenumi}{(\arabic{enumi})}
\item \textbf{Workflow value varies across formal targets and benchmarks.} Direct, the raw model baseline, leads the five-workflow SQL comparison in F1, while DRIFT improves both precision and recall on AmP. On AmP, DRIFT gains 15.98 F1 points at 2.03 times Direct's token cost; LoC uses 18.65 times Direct's tokens with lower F1.
\item \textbf{Explicit interpretation introduces a coverage--precision tradeoff.} Within AutoCedar, interpretation before generation lowers SQL F1 and increases AMBROSIA token use. On AMBROSIA, free prose recovers more reference readings while admitting more unmatched outputs.
\item \textbf{Review feedback provides modest improvements.} Adding Kimi at AutoCedar's review stage improves SQL F1, with K3 providing larger gains than K2.6.
\item \textbf{AutoCedar shows greater relative stability across input ambiguity.} Its smaller clear-to-ambiguous F1 decline suggests potential for handling ambiguity, while lower aggregate F1 leaves a net performance benefit unestablished.
\end{enumerate}
Together, these analyses help evaluate how autoformalization workflows handle the complex inputs expected in real-world operation.

\section{Related Work}
\paragraph{Ambiguity benchmarks and interpretation.}
AmbiQT studies lexical and structural ambiguity in SQL generation, AMBROSIA includes scope, attachment, and vagueness, and AmP provides logical forms for five linguistic ambiguity types \citep{bhaskar-etal-2023-benchmarking,saparina2024ambrosia,stengeleskin2024amp}. Saparina and Lapata generate preferred readings and use trained infilling to recover missing interpretations before SQL generation \citep{saparina-lapata-2025-disambiguate}. Their use of interpretation to improve coverage motivates a complementary question here: how do recovered readings affect precision and cost? Our free-interpretation condition increases AMBROSIA recall while lowering precision.

\paragraph{Intent validation and clarification.}
ARTEMIS translates requirements into structured natural language and uses user judgments of distinguishing traces to select temporal logic specifications \citep{mendoza2026artemis}. Its structured language stage informs our interpretation intervention. Monty combines conformance scores and code testing with user clarification when competing assertions remain plausible \citep{liu2026faithful}. These methods select intent with human input; our comparison evaluates the candidate sets that workflows produce autonomously.

\paragraph{Formalization workflows.}
Clover checks consistency among code, specifications, and descriptions \citep{sun2024clover}; Roundtrip uses backtranslation to identify and repair translation disagreements \citep{amrollahi2026roundtrip}. AutoCedar combines intent review with verifier-guided access-control policy synthesis \citep{vatsa2026autocedar}. LoC decomposes formalizations and alternates semantic and compiler feedback, while DRIFT retrieves dependencies and examples for theorem formalization \citep{shi2026loc,zhang2026drift}. We preserve these stage roles when adapting their formal targets, and evaluate how each complete workflow recovers multiple readings.

\paragraph{Feedback and evaluation.}
Self-Refine studies iterative model feedback, while Huang et al. find that intrinsic self-correction can reduce reasoning accuracy \citep{madaan2023selfrefine,huang2024selfcorrect}. Our review intervention varies the reviewer model at a fixed stage of AutoCedar. PICARD constrains SQL syntax, and test-suite evaluation assesses semantic correctness through execution \citep{scholak-etal-2021-picard,zhong-etal-2020-semantic}. We evaluate both the correctness of returned answers and the coverage of a benchmark's reference set, linking formal checks to ambiguity recovery.

\section{Autoformalization under Ambiguity}
\label{sec:method}
\subsection{Task and Available Information}
For an input $q$ and available context $D$, a workflow returns a set $C(q,D)$ of SQL queries or first-order logical forms. The benchmark provides a reference set $\mathcal{R}(q,D)$ representing its supported readings. A supported reading is an interpretation of $q$ grounded in $D$ and represented by a reference formalization; clear inputs have one reading, while ambiguous inputs have several. Evaluation groups equivalent formal outputs into answer classes before matching the returned and reference sets (Section~\ref{sec:scoring}). Reference outputs and reading counts enter only at evaluation.

The three datasets jointly evaluate ambiguity across inputs and formal targets. Each supplies a natural-language input and context for generation, with reference formalizations reserved for scoring (Supplementary Figure~\ref{fig:io}).

\paragraph{AMBROSIA.}
The input is a clear or ambiguous question paired with a database schema; the workflow returns SQL queries for its readings. For example, ``What size are all swimming pools and hot tubs in Miami'' permits restricting both facility types to Miami or only the hot tubs (Supplementary Table~\ref{tab:sample-evidence-3}). The reference queries encode these alternatives. Scoring executes predicted and reference SQL on the supplied database and matches their observed results, preserving required ordering.

\paragraph{AmbiQT.}
The input combines a question with a schema modified to introduce lexical or structural alternatives. The output is a set of SQL queries: in Figure~\ref{fig:running-example}, ``countries'' selects either \texttt{nationality} or \texttt{citizenship}. Scoring checks queries against the modified schema, applies the benchmark's mappings, and compares execution results on the underlying database (Supplementary Section~\ref{app:ambiqt-evaluation}).

\paragraph{AmP.}
The input is a sentence with a public predicate vocabulary and first-order formula grammar; the output is a set of logical forms for its readings. An illustrative sentence, ``Ada smiled,'' becomes $\exists a\,(\mathrm{smiled}(a)\land\mathrm{agent}(a,\mathrm{Ada}))$, with $a$ denoting an event. Ambiguous inputs yield alternatives in scope, attachment, or coreference. Scoring applies the same canonicalization to predictions and references and matches their canonical forms (Supplementary Section~\ref{app:extension}).

\subsection{Workflows and Adaptations}
\label{sec:architectures}
\label{sec:adaptations}
\begin{figure*}[!t]
\centering
\includegraphics[width=\textwidth]{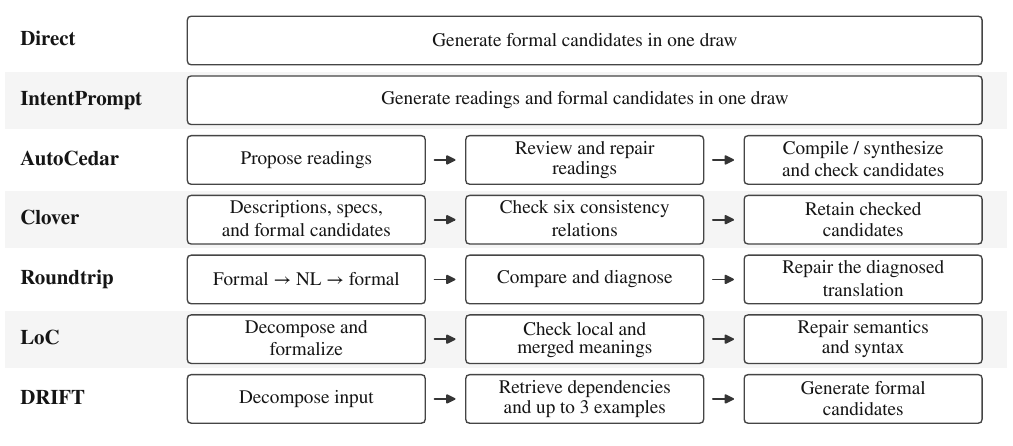}
\caption{Stages of the seven adapted workflows, read from left to right; NL denotes natural language. Model judgments inspect meaning, formal checks use execution or solvers, and retrieval supplies dependencies and examples. Each workflow returns a candidate set for external scoring. The AutoCedar interventions add interpretation before proposal generation or Kimi feedback at review. Section~\ref{sec:adaptations} describes the checks and bounded repairs.}
\label{fig:architecture}
\end{figure*}

We preserve each workflow's generation, review, repair, or retrieval stages while adapting its formal target and supporting checks. The seven workflows below follow the order in Figure~\ref{fig:architecture}; each returns a candidate set for external scoring.

\paragraph{Direct.}
The raw model generates formal candidates in one draw from the input, context, and target-language instructions, returning SQL queries or AmP formulas.

\paragraph{IntentPrompt.}
This interpretation-first prompting control requests paired natural-language readings and formal outputs through the source implementation's intent interface. It uses the shared generator and target-language instructions, adapted to SQL lists or AmP formulas.

\paragraph{AutoCedar.}
Originally designed for access-control policies with human intent review, AutoCedar combines target review with verifier-guided synthesis \citep{vatsa2026autocedar}. Our adaptations use model review: SQL deterministically compiles the accepted representation, with Cedar/cvc5 filter checks and SQLite syntax and schema checks; AmP uses Z3 to check property bounds and equivalence to the reviewed target. Supplementary Sections~\ref{app:extension} and~\ref{app:autocedar-sql} give representations and repair budgets.

\paragraph{Clover.}
Clover checks six directed consistency relations among descriptions, specifications, and code \citep{sun2024clover}. In SQL, four relations use execution and two use model judgments of descriptions; AmP substitutes quantified first-order equivalence checks for the formal relations. A candidate is retained when all six checks accept.

\paragraph{Roundtrip.}
Roundtrip backtranslates a formal output into natural language, formalizes that reconstruction, and diagnoses disagreements to repair the responsible translation \citep{amrollahi2026roundtrip}. SQL compares answer sets; AmP checks reconstructed formulas. Clover and Roundtrip use the supplied AMBROSIA databases and fixed public-schema test databases for internal AmbiQT checks (Supplementary Table~\ref{tab:adaptation-procedures}).

\paragraph{LoC.}
LoC decomposes formalization into nonrecursive SQL common table expressions or parameterized AmP definitions, then alternates semantic and syntax repair \citep{shi2026loc}. Local backtranslation omits the original question; model judgments check local purposes and combined meaning again after repair. Composition checks SQL dependencies and names or uses capture-avoiding substitution for AmP.

\paragraph{DRIFT.}
DRIFT decomposes the input, retrieves dependencies with a frozen encoder, and greedily selects up to three examples for final generation \citep{zhang2026drift}. Independently authored SQL or logical-form libraries replace the original mathematical material. Supplementary Section~\ref{app:extension} details the LoC and DRIFT procedures and retrieval libraries.

\subsection{Interpretation and Review Interventions}
\label{sec:interventions}
AutoCedar is the common pipeline for these two interventions because it exposes both an input position before generation and an intent-review stage. The seven-workflow comparisons evaluate each workflow separately; the interventions vary these two positions within AutoCedar (Figure~\ref{fig:architecture}).

\paragraph{Interpretation before generation.}
We ask whether making readings explicit improves reference coverage while retaining precision. Free interpretation supplies an ordinary prose account of candidate readings. Structured interpretation records meaning, source phrases, outputs, entities, conditions, and unresolved alternatives, inspired by ARTEMIS \citep{mendoza2026artemis}. Each is passed alongside the original question and context to AutoCedar, whose review, repair, and checking stages remain in place.

\paragraph{Different-family LLM reviewer.}
We ask whether feedback from a different model family improves the returned set. Kimi emulates a human response at AutoCedar's intent-review step, where the original architecture solicits human feedback \citep{vatsa2026autocedar}. The conditions share initial proposals and use no additional reviewer, Kimi K3, or Kimi K2.6. DeepSeek continues to generate and repair SQL.

\section{Experimental Design}
\label{sec:experiments}
\subsection{Research Questions and Comparisons}
We explore four research questions using the collected statistics:
\begin{description}
\item[RQ1:] \textbf{How does workflow choice affect performance across SQL and first-order logical forms?} All seven workflows share 1,000 AmP inputs. SQL compares five generation/checking workflows (Table~\ref{tab:main}) and, separately, LoC with DRIFT (Table~\ref{tab:sql-extension}).
\item[RQ2:] \textbf{How does interpretation before generation affect accuracy and token cost?} AutoCedar's base, free, and structured conditions share 381 scored AMBROSIA inputs; base and structured share 282 AmbiQT inputs (Table~\ref{tab:layers}).
\item[RQ3:] \textbf{How much does a different-family LLM reviewer improve the returned set?} AutoCedar's reviewer conditions share initial proposals and yield 3,432 scored pairs on AMBROSIA and 2,591 on AmbiQT (Table~\ref{tab:confirmation-paired}).
\item[RQ4:] \textbf{How does performance change from clear to ambiguous inputs, and how does this relate to aggregate performance?} AMBROSIA's source labels separate clear and ambiguous inputs within each comparison (Tables~\ref{tab:main} and~\ref{tab:sql-extension}).
\end{description}

\subsection{Evaluation}
\label{sec:scoring}
Let $r$ count reference answer classes, $c$ distinct valid predicted answer classes, and $i$ distinct invalid predictions after exact deduplication. If $m$ is the maximum number of one-to-one prediction--reference matches, then
\begin{equation}
P=\frac{m}{c+i},\qquad R=\frac{m}{r},\qquad F_1=\frac{2PR}{P+R}.
\end{equation}
\noindent Eligible cases with no predictions or no matches receive zero. All found measures complete reference coverage ($m=r$); Exact set additionally requires no extra predictions ($m=r=c+i$). In Figure~\ref{fig:running-example}, Direct matches both reference classes and has one additional prediction after grouping, yielding $P=2/3$, $R=1$, and $F_1=0.8$. Roundtrip matches exactly the two reference classes and receives one for each metric.

SQL queries with identical observed results form one answer class, so several query strings can represent the same predicted answer. A syntactically invalid query remains an invalid prediction and contributes to $i$. AMBROSIA scoring preserves required ordering; AmbiQT applies the benchmark schema mappings before comparing results (Supplementary Section~\ref{app:ambiqt-evaluation}). AmP applies the same canonicalization to predictions and references and matches their canonical forms. We macro-average the case metrics.

Unusable references and unavailable scores remain separate coverage categories, with each table reporting its numeric denominator. Supplementary Section~\ref{app:cohorts} and Table~\ref{tab:unscored-reasons} give eligibility and unscored reasons. Supplementary Figure~\ref{fig:venn} relates formal checks to supported-reading coverage, and Section~\ref{app:fullresults}, Table~\ref{tab:expanded-complete}, reports full answer-set recovery by condition.

\subsection{Implementation and Statistical Comparisons}
\label{sec:settings}
DeepSeek generates and repairs outputs through the recorded API alias \texttt{deepseek-flash}. The initial SQL and AmP/LoC/DRIFT request configurations enable thinking with low reasoning effort. Generation, repair, and any native review calls use DeepSeek; only the SQL feedback intervention adds Kimi K3 or Kimi K2.6. Each workflow retains its own bounded generation and repair schedule. DRIFT uses the frozen \texttt{BAAI/bge-small-en-v1.5} encoder and retrieves from independently authored SQL or AmP language examples. Supplementary Table~\ref{tab:settings} and Sections~\ref{app:extension} and~\ref{app:artifacts} provide request settings, prompt roles, and budgets.

Paired comparisons average per-input score differences. Interpretation intervals use database-cluster resampling; Holm correction covers the three AMBROSIA F1 contrasts. Interpretation token means use all 382 AMBROSIA inputs; the corresponding AmbiQT runs have different cost definitions. Both token analyses aggregate recorded usage across physical attempts. Interpretation uses its published per-input attempt totals; AmP and SQL LoC/DRIFT reconstruct prompt-plus-completion totals over the full saved generation, review, repair, and operational recovery histories. Physical attempts are counted once, with reasoning already included in completion. These comparisons use shared inputs with complete usage: 995 for AmP, 3,814 for AMBROSIA, and 3,044 for AmbiQT. Their full score cohorts remain unchanged. Supplementary Table~\ref{tab:token-coverage} gives cost coverage and components. Supplementary Section~\ref{app:statistics} and Table~\ref{tab:interpretation-intervals} give the existing resampling results for all six metrics. Means and differences are rounded independently from their full-precision values to two decimals.

Each workflow's 3,437 AMBROSIA inputs form a separately selected cohort; Direct and AutoCedar share 3,087 scored inputs. AmbiQT includes 2,741 for Direct, AutoCedar, and IntentPrompt and 3,046 for Clover and Roundtrip. LoC and DRIFT share a broader comparison of 3,819 AMBROSIA and 3,046 AmbiQT inputs. Each mean includes every numeric score in its published cohort; paired comparisons use shared scored inputs. Supplementary Tables~\ref{tab:cohorts} and~\ref{tab:extension-coverage} give the reference and output-level exclusions, and Table~\ref{tab:common-five} reports the five-workflow intersection.

AmP includes 1,000 shared inputs, with 200 from each of five ambiguity types: prepositional phrase attachment, quantifier scope, inverse scope, pronominal coreference, and conjunction. The interpretation comparison includes 382 AMBROSIA and 305 AmbiQT inputs, of which 381 and 282 have scores. Its AutoCedar baseline is labeled as the interpretation subset.

% Queue the cohort table before the results and keep table floats in numerical order.
\begin{table*}[!t]
\centering\small
\setlength{\tabcolsep}{4pt}
\caption{SQL generation and checking comparison. Means are percentages on each workflow's scored inputs. AMBROSIA's clear/ambiguous columns give scored counts and F1; all AmbiQT inputs are ambiguous. Direct leads overall F1 and precision in both datasets. Shared-input results and paired contrasts are in Supplementary Tables~\ref{tab:common-five} and~\ref{tab:paired}.}
\label{tab:main}
\begin{tabular}{@{}llrrrrrr@{}}
\toprule
Dataset & Workflow & Scored & F1 & Precision & Recall & Clear $N$ / F1 & Ambiguous $N$ / F1\\
\midrule
AMBROSIA & Direct & 3,431 & \textbf{53.35} & \textbf{51.19} & 62.29 & 2,400 / 58.36 & 1,031 / 41.68\\
 & IntentPrompt & 3,431 & 42.95 & 36.27 & 62.54 & 2,407 / 44.84 & 1,024 / 38.51\\
 & AutoCedar & 3,432 & 50.16 & 45.06 & \textbf{67.81} & 2,400 / 53.61 & 1,032 / 42.14\\
 & Clover & 3,430 & 41.30 & 35.82 & 57.97 & 2,402 / 43.90 & 1,028 / 35.23\\
 & Roundtrip & 3,433 & 38.65 & 35.50 & 47.48 & 2,409 / 42.26 & 1,024 / 30.16\\
\midrule
AmbiQT & Direct & 2,592 & \textbf{60.85} & \textbf{58.75} & \textbf{68.67} & --- & 2,592 / 60.85\\
 & IntentPrompt & 2,585 & 42.26 & 37.94 & 52.90 & --- & 2,585 / 42.26\\
 & AutoCedar & 2,591 & 54.45 & 50.89 & 65.69 & --- & 2,591 / 54.45\\
 & Clover & 2,871 & 20.69 & 19.19 & 26.56 & --- & 2,871 / 20.69\\
 & Roundtrip & 2,874 & 44.68 & 41.58 & 52.31 & --- & 2,874 / 44.68\\
\bottomrule
\end{tabular}
\end{table*}

\section{Results and Findings}
\subsection{Workflow Choice on SQL and Logical Forms}
\begin{figure*}[!t]
\centering
\includegraphics[width=0.96\textwidth]{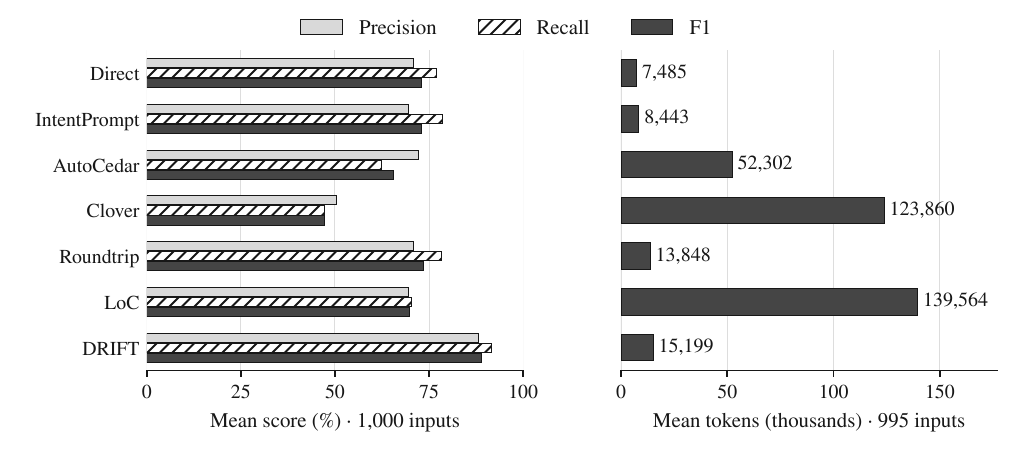}
\caption{AmP quality and inference cost. The left panel uses all 1,000 shared scored inputs; DRIFT's F1 advantage over Direct is 15.98 percentage points. The right panel sums saved physical-attempt tokens per input on the 995 inputs with complete usage for every workflow. Unknown costs affect the token panel only.}
\label{fig:amp-comparison}
\end{figure*}

On the same 1,000 AmP inputs, DRIFT reaches 89.02\% F1, compared with 73.05\% for Direct and 73.51\% for Roundtrip.
DRIFT improves both precision and recall (Figure~\ref{fig:amp-comparison}); Supplementary Table~\ref{tab:amp} gives the means. Its generation step combines the original input with retrieved dependencies and up to three examples from an independently authored library of 11 AmP language examples. On the 995 inputs with complete costs across all seven workflows, Direct uses 7,485 tokens per input and DRIFT uses 15,199, a 2.03-fold increase. AutoCedar, Clover, and LoC use 52,302, 123,860, and 139,564 tokens respectively, with lower F1 than Direct on the full scored set.

On SQL, Direct has the highest overall F1 and precision among the five workflows in Table~\ref{tab:main}. AutoCedar's higher AMBROSIA recall is offset by its lower precision, while on AmbiQT both metrics fall below Direct (Figure~\ref{fig:sql-comparison}). The same pattern holds on all five methods' shared scored inputs in Supplementary Table~\ref{tab:common-five}, and paired comparisons with Direct favor Direct overall (Supplementary Table~\ref{tab:paired}).

\begin{figure*}[!t]
\centering
\includegraphics[width=\textwidth]{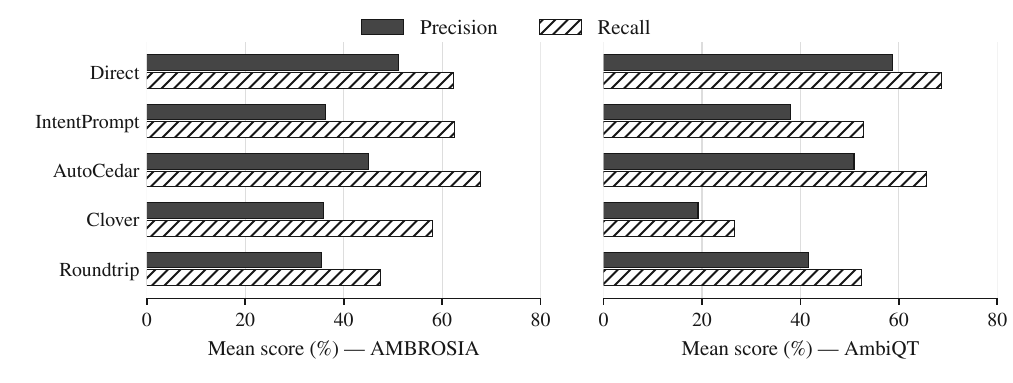}
\caption{SQL precision and recall on each workflow's scored cohort: 3,430--3,433 AMBROSIA inputs and 2,585--2,874 AmbiQT inputs (Table~\ref{tab:main}). On AMBROSIA, AutoCedar's additional coverage comes with lower precision. On AmbiQT, Direct leads both metrics. Grouped bars show precision and recall in the same workflow order; both dataset panels use the same scale.}
\label{fig:sql-comparison}
\end{figure*}

Clover's AmbiQT means are low in both precision and recall. A saved example makes the distinction between its consistency checks and benchmark recovery concrete: for the input in Figure~\ref{fig:running-example}, Clover returns nationality and citizenship together in one query. Its reconstruction checks match, but the query matches neither reference answer. Supplementary Section~\ref{app:samples}, Tables~\ref{tab:sample-evidence-1}--\ref{tab:sample-evidence-10}, provides the outputs and intermediate representations. This case illustrates how internally matching representations can preserve an incorrect candidate set.

In the separate decomposition comparison, LoC and DRIFT have similar AMBROSIA F1, including on ambiguous questions. DRIFT leads LoC by 8.20 points on their shared scored AmbiQT inputs (Table~\ref{tab:sql-extension}). The comparison evaluates retrieval with illustration against modular repair on the same inputs. The complete-cost comparison also favors DRIFT: its mean token use is 80.7\% below LoC on AMBROSIA and 69.5\% below it on AmbiQT.

\begin{table*}[!t]
\begin{minipage}[t]{0.48\textwidth}
\vspace{0pt}
\centering\small
\setlength{\tabcolsep}{3pt}
\begin{minipage}[t][4\baselineskip][t]{\linewidth}
\caption{SQL decomposition comparison. F1 uses shared scored inputs; tokens use shared inputs with complete usage (Supplementary Tables~\ref{tab:extension-coverage} and~\ref{tab:token-coverage}).}
\label{tab:sql-extension}
\end{minipage}\par
\begin{tabular*}{\linewidth}{@{\extracolsep{\fill}}llrrr@{}}
\toprule
Dataset & Group & $N$ & LoC & DRIFT\\
\midrule
\multicolumn{5}{l}{F1 (\%)}\\
AMBROSIA & All & 3,813 & 53.34 & 52.75\\
 & Clear & 2,666 & 58.88 & 57.86\\
 & Ambiguous & 1,147 & 40.48 & 40.86\\
AmbiQT & Ambiguous & 2,884 & 38.98 & 47.17\\
\midrule
\multicolumn{5}{l}{Mean tokens/input}\\
AMBROSIA & All & 3,814 & 31,211 & 6,015\\
AmbiQT & All & 3,044 & 19,698 & 6,003\\
\bottomrule
\end{tabular*}
\end{minipage}
\hfill
\begin{minipage}[t]{0.48\textwidth}
\vspace{0pt}
\centering\small
\setlength{\tabcolsep}{3pt}
\begin{minipage}[t][4\baselineskip][t]{\linewidth}
\caption{AutoCedar interpretation comparison. Scores (\%) use 381 shared AMBROSIA and 282 AmbiQT inputs; tokens use 382 AMBROSIA inputs (Supplementary Tables~\ref{tab:interpretation-means} and~\ref{tab:interpretation-intervals}).}
\label{tab:layers}
\end{minipage}\par
\begin{tabular*}{\linewidth}{@{\extracolsep{\fill}}lrrrr@{}}
\toprule
Condition & F1 & P & R & Tokens\\
\midrule
\multicolumn{5}{l}{AMBROSIA}\\
Base & \textbf{50.05} & \textbf{44.41} & 67.80 & 45,538\\
+ Free & 43.24 & 34.70 & \textbf{72.75} & 154,105\\
+ Structured & 43.61 & 37.15 & 65.92 & 99,431\\
\midrule
\multicolumn{5}{l}{AmbiQT}\\
Base & \textbf{51.67} & \textbf{48.33} & \textbf{62.23} & ---\\
+ Structured & 44.72 & 39.49 & 59.75 & ---\\
\bottomrule
\end{tabular*}
\end{minipage}
\end{table*}

\subsection{Interpretation before Generation}
\label{sec:interpretation-results}

Structured interpretation lowers F1 by 6.44 points on AMBROSIA and 6.95 on AmbiQT, with both paired intervals entirely below zero (Supplementary Table~\ref{tab:interpretation-intervals}); Table~\ref{tab:layers} gives the condition means.
Free interpretation raises AMBROSIA recall but reduces precision enough to lower F1. Thus, its additional reference coverage accompanies a less precise returned set.

The saved output counts clarify this tradeoff. Across the shared scored AMBROSIA inputs, AutoCedar returns 990 distinct answers, free interpretation 1,491, and structured interpretation 1,250 (Supplementary Table~\ref{tab:candidate-counts}). Structured interpretation yields a smaller set and higher precision than free prose, but also lower recall; their paired F1 interval includes zero. Supplementary Tables~\ref{tab:sample-evidence-11}--\ref{tab:sample-evidence-20} show the supplied interpretations and returned SQL.

Both added interpretation stages also increase mean token use: about 3.4 times the base for free prose and 2.2 times for structured interpretation. The cost and F1 values appear together in Table~\ref{tab:layers}. This comparison favors the base pipeline when choosing among these three interpretation conditions.

\subsection{Different-family LLM Review}
\label{sec:feedback-results}
Kimi K3 improves paired F1 by 1.93 points on AMBROSIA and 2.04 on AmbiQT (Table~\ref{tab:confirmation-paired}). Kimi K2.6 gives smaller gains. The direct K3--K2.6 comparison favors K3 on both datasets, with both 95\% database-cluster intervals above zero (Supplementary Table~\ref{tab:reviewer-direct-intervals}).

\begin{table}[!t]
\centering\small
\caption{Review feedback yields modest paired gains. Each contrast uses inputs scored with the named reviewer and with no additional reviewer. Differences are percentage points.}
\label{tab:confirmation-paired}
\begin{tabular}{@{}llrr@{}}
\toprule
Dataset & Reviewer & Pairs & $\Delta$F1\\
\midrule
AMBROSIA & Kimi K3 & 3,432 & $+1.93$\\
 & Kimi K2.6 & 3,432 & $+1.23$\\
AmbiQT & Kimi K3 & 2,591 & $+2.04$\\
 & Kimi K2.6 & 2,591 & $+0.24$\\
\bottomrule
\end{tabular}
\end{table}

Supplementary Table~\ref{tab:confirmation} shows the precision and recall changes under each reviewer. On AmbiQT, K3 raises precision from 50.89\% to 54.42\% while recall changes from 65.69\% to 65.34\%. K2.6 reaches 52.02\% precision and 64.61\% recall, producing a smaller F1 gain. K3 combines the larger precision gain with the smaller recall loss. Supplementary Figure~\ref{fig:bar} visualizes the paired gains. The initial confirmation study is retained separately in Supplementary Table~\ref{tab:complete}, with examples in Tables~\ref{tab:sample-evidence-21}--\ref{tab:sample-evidence-30}.

Fully automated review still leaves substantial errors in these SQL experiments. AutoCedar's original architecture uses human intent review to establish the target for formal verification \citep{vatsa2026autocedar}. Human clarification serves this crucial role when deployment requires selecting a user's intended action from competing readings \citep{mendoza2026artemis}.

\subsection{Clear and Ambiguous Questions}
\label{sec:strata-results}
AutoCedar's mean F1 falls from 53.61\% on clear AMBROSIA inputs to 42.14\% on ambiguous inputs. Its 11.46-point drop is smaller than Direct's 16.68-point drop (Supplementary Table~\ref{tab:interpretation-groups}).
Clear questions form about 70\% of these AMBROSIA cohorts. AutoCedar's overall F1 is 50.16\%, compared with 53.35\% for Direct (Table~\ref{tab:main}).

On the 3,087 inputs scored by both workflows, AutoCedar is 3.45 points below Direct overall (Supplementary Table~\ref{tab:paired}). Its paired deficit is 5.02 points on clear inputs, whereas its 0.18-point advantage on ambiguous inputs leaves the means numerically close (Supplementary Table~\ref{tab:strata}). The smaller decline indicates greater relative stability across clear and ambiguous inputs, suggesting potential for workflow design to improve ambiguity handling. Its lower clear-input baseline and aggregate F1, however, leave a net performance benefit unestablished. These results therefore do not establish that autoformalization inherently improves ambiguity handling.

\section{Discussion and Scope}
\subsection{Implications for Agentic Systems}
\paragraph{Preserve intent through execution.}
An agent's generated SQL query determines which records are processed downstream, while a logical form makes the chosen reading explicit. Our precision--coverage results make candidate acceptance a central design choice: generating more interpretations can recover missed readings while introducing additional incorrect candidates. A system should establish the intended scope before executing a sensitive operation and retain the relationship between that scope and the artifact it executes. Related access-control work places intent validation before policy verification, providing a complementary application of this design principle \citep{cutler2024cedar,vatsa2026autocedar}.

\paragraph{Design stages around the returned set.}
The free-interpretation stage recovers more AMBROSIA reference answers, but also admits enough unmatched outputs to lower precision and F1. The ARTEMIS-inspired structured layer reduces this expansion relative to free prose, yet both conditions remain below the base pipeline. These results connect interpretation-first coverage goals \citep{saparina-lapata-2025-disambiguate} to the acceptance decisions made by later stages. Representation, placement, and candidate selection must work together; additional stages alone do not resolve the precision--coverage tradeoff.

\paragraph{Match the workflow to its task.}
Direct leads the five-workflow comparison on both SQL benchmarks, while DRIFT leads on AmP. Workflow rankings therefore vary across these benchmark and formal-target settings. A system designer should evaluate a workflow on the intended language, task, and input distribution. Clear inputs also matter: AutoCedar's smaller clear-to-ambiguous F1 decline indicates greater relative stability, but its lower clear-input accuracy leaves aggregate F1 below Direct. Ambiguity robustness and overall accuracy are separate deployment requirements.

\paragraph{Use review to support intent decisions.}
AutoCedar's original design uses human approval to establish the intended target \citep{vatsa2026autocedar}; our adaptation places model review at this stage. Roundtrip automates diagnosis, and its cross-model experiment improves repair by changing the diagnosing model \citep{amrollahi2026roundtrip}. The modest gains from Kimi show the value of reviewer choice within our AutoCedar comparison. Human clarification remains important when a system must select a user's intended action from several plausible readings.

\paragraph{Require a measurable return on inference cost.}
On the AMBROSIA interpretation subset, the base pipeline has both the highest F1 and the lowest mean token use. Structured and free interpretation consume 2.18 and 3.38 times its tokens while reducing F1. This is a direct cost--performance regression for the tested additions. The AmP comparison gives a different tradeoff: DRIFT improves F1 while using 2.03 times Direct's tokens, whereas LoC uses 18.65 times its tokens with lower F1. Deployment budgets should account for every generation, review, and repair call and evaluate the benefit of the complete workflow. For routine inputs, an extra stage should preserve answer quality and justify its cost even when the final answer stays the same.

\subsection{Limitations}
The experiments use one generator family and bounded adaptations of the source workflows. Their request counts and repair schedules differ. The SQL workflow groups have different fixed cohorts; paired contrasts and common-input results address comparisons within each group. The interpretation experiments use smaller fixed subsets, and reviewer feedback is evaluated on SQL. SQL correctness uses observed execution results, while AmP uses the recorded canonical-form matching protocol. Cost comparisons cover the AMBROSIA interpretation study, all seven AmP workflows, and SQL LoC/DRIFT. They measure observed token use through the accepted output, including saved operational recovery attempts; twelve extension inputs have incomplete usage and retain unknown total costs.

\section{Conclusion}
By adapting seven autoformalization workflows to ambiguity benchmarks, we observe performance regressions in the SQL comparisons and gains on logical forms. Direct leads the five-workflow SQL comparison, while DRIFT improves AmP precision and recall. Within AutoCedar, the ARTEMIS-inspired structured interpretation layer lowers SQL F1, showing that ambiguity-oriented components need to be integrated through careful workflow design. Different-family model review yields modest gains, and AutoCedar's smaller clear-to-ambiguous decline suggests relative stability while leaving a net performance benefit unestablished.

For agentic systems that query databases or construct logical forms, these findings connect workflow design to the accuracy and cost of formalized outputs. Recovering plausible readings, selecting the intended one, and preserving it through formalization are joint requirements for reliable execution. The measured cost regressions also matter at scale: prior interpretation uses 2.2--3.4 times as many tokens on the AMBROSIA subset while reducing F1. Repeating such overhead across large request volumes increases resource demand without a corresponding quality gain in this comparison. Effective deployment therefore requires workflows that preserve accuracy on routine inputs, handle ambiguous intent explicitly, and justify the cumulative cost of generation, checking, and repair.
\label{mainend}

\section*{Ethical Considerations}
The study uses existing benchmark inputs and saved model outputs, without recruiting human participants. SQL evaluation does not modify the databases. Reuse of data and upstream code follows their source licenses.

\section*{AI Usage Disclosure}
OpenAI's Codex assistant supported literature inspection, experiment implementation and debugging, analysis scripts, manuscript revision, and document checks. DeepSeek generated and repaired the experimental SQL queries and logical forms; Kimi K3 and Kimi K2.6 provided the specified review feedback. The authors directed the study, reviewed the data and manuscript, and approved the final scientific claims.

\clearpage
\setcounter{table}{0}
\setcounter{figure}{0}
\renewcommand{\thetable}{S\arabic{table}}
\renewcommand{\thefigure}{S\arabic{figure}}
\makeatletter
\setlength{\@fptop}{0pt}
\setlength{\@fpsep}{14pt}
\setlength{\@fpbot}{0pt plus 1fil}
\setlength{\@dblfptop}{0pt}
\setlength{\@dblfpsep}{14pt}
\setlength{\@dblfpbot}{0pt plus 1fil}
\makeatother
\twocolumn[
\begin{center}
{\Large\bfseries Supplementary Material\par}
\vspace{0.5em}
{\large Autoformalization for Ambiguous Inputs\par}
\end{center}
\vspace{0.5em}
]
\appendix
\section{Workflow Adaptation Protocols}
\label{app:extension}
\paragraph{Input sets and labels.}
All seven workflows use the same AmP inputs, drawn from the five source zero-shot ambiguous test sets. The selected set contains 200 inputs per type, for 1,000 inputs in total. LoC and DRIFT also share their inputs and reference exclusions on each SQL benchmark. The comparisons retain the source ambiguity labels. Table~\ref{tab:extension-coverage} reports scoring coverage, and Main paper Table~\ref{tab:sql-extension} separates clear and ambiguous SQL inputs.

\begin{table*}[t]
\centering\small
\caption{Scoring coverage for the AmP and SQL decomposition comparisons. Each record is one workflow on one input. The 336 unscored SQL records correspond to 168 inputs whose original reference sets could not be scored, shared by LoC and DRIFT.}
\label{tab:extension-coverage}
\begin{tabular}{@{}llrrr@{}}
\toprule
Dataset & Workflows & Records & Scored & Not scored\\
\midrule
AmP & All seven & 7,000 & 7,000 & 0\\
AMBROSIA & LoC and DRIFT & 7,638 & 7,626 & 12\\
AmbiQT & LoC and DRIFT & 6,092 & 5,768 & 324\\
\midrule
Total & & 20,730 & 20,394 & 336\\
\bottomrule
\end{tabular}
\end{table*}

\paragraph{AmP adaptation.}
Direct returns formulas in one draw. IntentPrompt uses an interpretation first prompting interface. AutoCedar proposes source-grounded properties, checks floor and ceiling constraints, reviews the target, and realizes it through bounded counterexample-guided synthesis. The AmP implementation uses the specified predicate language and Z3, with two target repair rounds and three synthesis iterations. Clover checks six directed consistency relations among descriptions, specifications, and formal outputs. Four formal relations use quantified FOL SMT equivalence. Two description relations use source only reconstruction and model judgments. It exports an initial formula when all six checks accept. Roundtrip reconstructs descriptions from formulas, formalizes them again, and repairs the diagnosed translation stage. It allows five encoding attempts and three semantic repairs, returning the latest first stage set.

\paragraph{LoC.}
Both adaptations retain modular formalization, local backtranslation, local purpose checks, merged reconstruction, three level semantic judgment, discrepancy rechecking, and alternating semantic and syntax repair. Backtranslation sees the formal artifact and public context, without the intended question or module purpose. Semantic judgments are recomputed after syntax repair. The bounds are five outer rounds, two semantic repairs, and three syntax repairs. AmP modules use capture-avoiding substitution. SQL modules use nonrecursive common table expressions with dependency and name capture checks. SQLite compilation checks syntax against the public schema. External scoring receives the latest complete candidate set, with internal acceptance recorded separately.

\paragraph{DRIFT.}
The adaptation follows the published decomposition, retrieval, illustration, and formalization procedure. A frozen \texttt{BAAI/bge-small-en-v1.5} encoder retrieves the top dependency by cosine similarity for each concept, followed by deduplication. Greedy example selection maximizes marginal dependency coverage, using semantic similarity to break ties and a budget of three examples. The final generation step receives the original input, selected dependencies and examples, and the context available during generation. SQL and FOL material replace the original mathematical library.
The SQL library contains 24 dependencies and 20 independently authored language examples. The AmP library contains 13 base dependencies and 11 examples authored from the public formula grammar and vocabulary, supplemented by the actor-visible predicate and constant definitions. These libraries were constructed independently of benchmark questions and reference outputs. DRIFT adds up to three retrieved examples to the shared task and language instructions; Direct uses its task prompt in one generation call. The frozen encoder revision is \texttt{5c38ec7c405ec4b44b94\allowbreak{}cc5a9bb96e735b38267a}.

\paragraph{External scoring and provenance.}
AmP uses the recorded \texttt{amp\_canonical\_set\_compat001} protocol. Both predictions and original references are canonicalized by the same procedure. Exact duplicate predictions are removed; valid predictions are grouped by canonical form, and distinct invalid predictions count against precision. SQL uses the strict observation and one-to-one matching evaluator, preserving typed values, bag or order requirements, native AmbiQT mappings, and invalid output penalties. Cases with an unknown checker outcome have no score. Corrections to SQL text decoding preserve the original references and saved predictions.

 AmP AutoCedar, Clover, LoC, and DRIFT use the revised stage prompts. Direct, IntentPrompt, and Roundtrip use the baseline generation protocols with compatible external scoring. All seven have 1,000 scored inputs. Their selected inputs and reference exclusions are shared. The AmP and SQL decomposition comparisons use DeepSeek for generation and any native feedback stages, with no human confirmation.

\section{Experimental Settings and Adaptations}
\label{app:artifacts}
\paragraph{Data and code availability.}
Code, model outputs, and per case scores are stored in the project repository. The analyses use the final cleaned release, including the corrected expanded SQL records and the AmP and SQL decomposition results, together with their recorded implementation snapshots.

% BEGIN APPENDIX SETTINGS

\subsection{Runtime and Model Settings}
Table~\ref{tab:settings} summarizes the inference and evaluation settings.

\begin{table*}[!t]
\centering\small
\caption{Recorded settings for the initial SQL and interpretation studies and the AmP and SQL decomposition extensions.}
\label{tab:settings}
\begin{tabular}{@{}p{0.23\textwidth}p{0.73\textwidth}@{}}
\toprule
Setting & Recorded value\\
\midrule
Local environment & Python on Linux, Cedar, cvc5 Linux ARM64, and SQLite. One thread per process.\\
Generation model and API & DeepSeek, recorded under \texttt{deepseek-flash}, accessed through streaming chat completion requests to \texttt{api.deepseek.com}. \\
Initial SQL request settings & Configured output limit of 32,768 tokens, thinking enabled, low reasoning effort, timeout of 600 seconds, and two overload retries. Provider default temperature. \\
Confirmation models & Kimi K3 or Kimi K2.6 supplies feedback. DeepSeek generates and repairs SQL.\\
AutoCedar budget & Up to two interpretation repairs per current property and one deterministic compilation candidate.\\
Clover and Roundtrip & Clover allows one retry after reconstruction execution failure. Roundtrip allows five initial encoding validation attempts per translation and three semantic repairs. \\
Evaluation limits & Read only SQLite, five seconds per query, 100,000 rows, and a progress check every 1,000 instructions. \\
Extension request settings & AmP and SQL LoC/DRIFT use a configured output limit of 393,216 tokens, a 1,800-second inference timeout, and a 20-second connection timeout. Thinking is enabled with low reasoning effort. \\
AmP request and solver limits & Up to four physical attempts per logical request, a shared limit of 64 concurrent requests, and a 10-second solver timeout. \\
\bottomrule
\end{tabular}
\end{table*}

\subsection{SQL Adaptation Procedures}
\label{app:adaptation}
Main paper Figure~\ref{fig:architecture} shows where each adaptation interprets, checks, and revises the input. Table~\ref{tab:adaptation-procedures} gives pseudocode for the generation procedures. 

\begin{table*}[t]
\centering\footnotesize
\caption{SQL adaptation procedures. The interpretation layer precedes AutoCedar, while simulated feedback acts inside its review process.}
\label{tab:adaptation-procedures}
\begin{tabular}{@{}p{0.18\textwidth}p{0.77\textwidth}@{}}
\toprule
Procedure & Operations for question $q$ and database context $D$\\
\midrule
Direct & Generate a SQL list from the benchmark prompt and parse the returned candidates.\\
AutoCedar & Enumerate readings of $q$. For each reading, review its agreement with $q$, allowing two repairs. Compile its SQL representation deterministically, check the fixed target, and collect the exported SQL candidates.\\
Roundtrip & Generate SQL $T_1$. Reconstruct language $T_2$ from $T_1$ and $D$. Generate SQL $T_3$ from $T_2$ and $D$. Compare the answer sets of $T_1$ and $T_3$. On disagreement, diagnose and repair the translation stage, up to three times. Return the latest $T_1$.\\
Clover & Generate a specification and SQL for each reading. Reconstruct their descriptions and check agreement in six directions, four using SQL results and two using text comparisons. Retry once after a reconstruction execution error. Return retained SQL.\\
IntentPrompt & Prompt for explicit readings and their SQL. Extract the interpretation and SQL fields from the returned XML.\\
Interpretation layer & Produce free prose or structured natural language. Pass it with $q$ and $D$ to AutoCedar.\\
Simulated feedback & Start from the saved shared proposal. At the confirmation step of AutoCedar, use no additional reviewer, a Kimi K3 reviewer, or a Kimi K2.6 reviewer. Continue the native review, repair, and target checks. \\
\bottomrule
\end{tabular}
\end{table*}

\subsection{Evaluation Input Sets and Coverage}
\label{app:cohorts}

\begin{table*}[tp]
\centering\small
\setlength{\tabcolsep}{4pt}
\caption{Counts are question and condition records. Not scored includes ineligible references and other cases without a score.}
\label{tab:cohorts}
\begin{tabular}{@{}lrrrrrr@{}}
\toprule
 & \multicolumn{3}{c}{AMBROSIA} & \multicolumn{3}{c}{AmbiQT}\\
\cmidrule(lr){2-4}\cmidrule(l){5-7}
Workflow & Total & \textbf{Scored} & Not scored & Total & \textbf{Scored} & Not scored\\
\midrule
\multicolumn{7}{l}{\textit{Expanded workflows and confirmation}}\\
Direct & 3,437 & 3,431 & 6 & 2,741 & 2,592 & 149\\
AutoCedar & 3,437 & 3,432 & 5 & 2,741 & 2,591 & 150\\
Roundtrip & 3,437 & 3,433 & 4 & 3,046 & 2,874 & 172\\
Clover & 3,437 & 3,430 & 7 & 3,046 & 2,871 & 175\\
IntentPrompt & 3,437 & 3,431 & 6 & 2,741 & 2,585 & 156\\
AutoCedar + Kimi K3 & 3,437 & 3,432 & 5 & 2,741 & 2,591 & 150\\
AutoCedar + Kimi K2.6 & 3,437 & 3,432 & 5 & 2,741 & 2,591 & 150\\
Expanded total & 24,059 & 24,021 & 38 & 19,797 & 18,695 & 1,102\\
\midrule
\multicolumn{7}{l}{\textit{Initial confirmation}}\\
No extra model & 382 & 381 & 1 & 305 & 282 & 23\\
Kimi K3 & 382 & 381 & 1 & 305 & 282 & 23\\
Kimi K2.6 & 382 & 380 & 2 & 305 & 281 & 24\\
\bottomrule
\end{tabular}

\vspace{6pt}
\begin{tabular}{@{}lrrr@{}}
\toprule
\textit{Prior interpretation on AMBROSIA} & & &\\
Workflow & Total & \textbf{Scored} & Not scored\\
\midrule
AutoCedar & 382 & 381 & 1\\
AutoCedar + free interpretation & 382 & 381 & 1\\
AutoCedar + structured interpretation & 382 & 381 & 1\\
\midrule
\textit{Prior interpretation on AmbiQT} & & &\\
Workflow & Total & \textbf{Scored} & Not scored\\
\midrule
AutoCedar & 305 & 282 & 23\\
AutoCedar + structured interpretation & 305 & 282 & 23\\
\bottomrule
\end{tabular}

\end{table*}

 Table~\ref{tab:expanded-complete} reports coverage for each condition. Table~\ref{tab:unscored-reasons} separates reference problems from other unscorable cases.

Eligibility is checked by input identity within each published comparison. Across the five SQL workflows, every shared input labeled reference-unscorable has no numeric score in the other workflows. Two Clover records use an output-level label where another workflow uses a reference-level label; both remain unscored. The original SQL and later LoC/DRIFT releases also differ in reference availability for ten AmbiQT inputs: 69 original workflow--condition records retain the reference-unavailable label, while LoC and DRIFT have numeric scores for those inputs. Each comparison uses its published scores and coverage.

\begin{table}[t]
\centering\small
\caption{Recorded reasons for unscored records in the expanded evaluation.}
\label{tab:unscored-reasons}
\begin{tabular}{@{}p{6.2cm}r@{}}
\toprule
Reason & Records\\
\midrule
\multicolumn{2}{l}{\textit{Reference problems}}\\
Reference unavailable & 69\\
Invalid reference & 977\\
No valid reference branch & 51\\
Reference execution failure & 37\\
\midrule
\multicolumn{2}{l}{\textit{Other unscored cases}}\\
Nonidentical SQL; empty-fixture evidence insufficient & 5\\
Tied \texttt{ORDER BY} values cut by \texttt{LIMIT}; result not uniquely specified & 1\\
\midrule
Total not scored & 1,140\\
\bottomrule
\end{tabular}
\end{table}

The AMBROSIA interpretation study includes both clear and ambiguous questions. Its baseline was generated alongside the two interpretation variants. The AmbiQT comparison reuses the initial AutoCedar baseline. Table~\ref{tab:cohorts} reports the selected and scored input counts.

The initial confirmation study compares no extra model with K3 and K2.6 on shared initial proposals. Table~\ref{tab:cohorts} reports coverage for each condition. 

\subsection{Adapting AutoCedar to SQL}
\label{app:autocedar-sql}
The SQL backend fixes the relational structure and predicate atoms. Cedar and cvc5 verify the filter structure formed by \texttt{AND}, \texttt{OR}, and \texttt{NOT} under the SQL values true, false, and unknown. SQL is exported when its deterministic Cedar translation matches the checked representation. SQLite \texttt{EXPLAIN} separately checks syntax and schema compatibility without executing the query.

\subsection{Preparing AmbiQT Outputs for Evaluation}
\label{app:ambiqt-evaluation}
Method specific parsers extract SQL candidate lists from model responses. During generation, the public schema becomes an empty SQLite database with quoted identifiers and \texttt{TEXT} columns for syntax and name checks.

During evaluation, candidates compile against the altered schema before the AmbiQT mappings for column names, table names, split tables, or aggregated tables translate them for execution on the original Spider database. The upstream rule suppresses the second branch for table splitting and aggregation when the first already matches. Scoring compares complete mapped results and groups reference slots with overlapping supported branches into one target. Private maps, reference SQL, and evaluation rows remain on the evaluation side.
\paragraph{AmbiQT consistency checks in the expanded evaluation.}
Clover and Roundtrip use a fixed bank of test databases derived from the public schema. The bank is independent of the candidate SQL and remains fixed throughout checking. The initial workflow study evaluates Clover and Roundtrip on AMBROSIA.

\paragraph{Saved output scoring.}
The expanded snapshot preserves returned SQL strings and uses the existing reference eligibility policy. It incorporates completed scores after checker repairs and records parser adaptations for 23 IntentPrompt outputs. All reported means are recomputed from the published per case scores.
% END APPENDIX SETTINGS
\paragraph{Candidate counts.}
Both interpretation variants increase the counts of returned SQL candidates and distinct answers after grouping queries with equivalent results. Table~\ref{tab:candidate-counts} reports the counts.

\begin{table}[t]
\centering\small
\caption{Candidate counts on shared scored AMBROSIA inputs. Free and structured are interpretation steps before AutoCedar. Distinct answers group equivalent SQL results.}
\label{tab:candidate-counts}
\begin{tabular}{@{}lrr@{}}
\toprule
Workflow & SQL candidates & Distinct answers\\
\midrule
AutoCedar & 1,664 & 990\\
Free interpretation & 3,353 & 1,491\\
Structured interpretation & 2,595 & 1,250\\
\bottomrule
\end{tabular}
\end{table}

\begin{table*}[!t]
\centering\small
\setlength{\tabcolsep}{5pt}
\caption{Full condition means for interpretation before AutoCedar: 381 shared scored AMBROSIA inputs and 282 AmbiQT inputs. Base denotes AutoCedar on that subset. Metrics are percentages. Mean tokens cover all 382 AMBROSIA inputs; the AmbiQT cost comparison has different recorded definitions. Paired differences and 95\% database-cluster intervals are in Supplementary Table~\ref{tab:interpretation-intervals}.}
\label{tab:interpretation-means}
\begin{tabular}{@{}llrrrrr@{}}
\toprule
Dataset & Interpretation subset condition & F1 & Precision & Recall & All found & Mean tokens\\
\midrule
AMBROSIA & Base AutoCedar & \textbf{50.05} & \textbf{44.41} & 67.80 & 58.53 & 45,538\\
 & + Free interpretation & 43.24 & 34.70 & \textbf{72.75} & \textbf{64.04} & 154,105\\
 & + Structured interpretation & 43.61 & 37.15 & 65.92 & 57.22 & 99,431\\
\midrule
AmbiQT & Base AutoCedar & \textbf{51.67} & \textbf{48.33} & \textbf{62.23} & \textbf{50.71} & ---\\
 & + Structured interpretation & 44.72 & 39.49 & 59.75 & 50.00 & ---\\
\bottomrule
\end{tabular}
\end{table*}

\subsection{Inference Cost Accounting}
\label{app:token-accounting}
We aggregate the provider's prompt and completion counters over each published generation history, including review, repair, and recorded recovery attempts. Completion counts already include reasoning tokens. Each physical usage receipt is counted once by its path and content hash; replay of a saved output adds no model call. The audit authenticates 216,856 receipts across 20,730 workflow--input records. Twelve records contain an attempt whose usage was not returned. Their known subtotals are retained, and their total costs remain unknown.

Table~\ref{tab:token-coverage} reports cost completeness before taking a common-input intersection. AmP has 995 inputs with complete usage for all seven workflows; LoC and DRIFT share 3,814 complete-cost AMBROSIA inputs and 3,044 AmbiQT inputs. These cost exclusions leave the numeric-score cohorts unchanged. The AMBROSIA interpretation study sums its published per-input physical-attempt counters (\texttt{total\_tokens\_known\_sum}); all 382 inputs in each condition have complete recorded usage. Both studies aggregate recorded attempt tokens, with the interpretation totals taken from that study and the extension totals reconstructed from its full saved histories. Each cost ratio uses the shared complete-cost cohort within its study (Main paper Table~\ref{tab:layers}; Table~\ref{tab:token-coverage}).

\begin{table*}[!ht]
\centering\small
\caption{Provider-token coverage and means for complete generation histories. Cost-complete counts are per workflow; common $N$ is the intersection used for the displayed prompt, completion, and total means. Counts include saved recovery attempts.}
\label{tab:token-coverage}
\begin{tabular}{@{}llrrrrrr@{}}
\toprule
Dataset & Workflow & Inputs & Cost-complete & Common $N$ & Prompt & Completion & Total\\
\midrule
AmP & Direct & 1,000 & 1,000 & 995 & 2,143 & 5,342 & 7,485\\
 & IntentPrompt & 1,000 & 1,000 & 995 & 2,466 & 5,977 & 8,443\\
 & AutoCedar & 1,000 & 999 & 995 & 32,128 & 20,175 & 52,302\\
 & Clover & 1,000 & 996 & 995 & 86,194 & 37,667 & 123,860\\
 & Roundtrip & 1,000 & 1,000 & 995 & 6,920 & 6,928 & 13,848\\
 & LoC & 1,000 & 1,000 & 995 & 101,191 & 38,373 & 139,564\\
 & DRIFT & 1,000 & 1,000 & 995 & 10,517 & 4,682 & 15,199\\
\midrule
AMBROSIA & LoC & 3,819 & 3,817 & 3,814 & 12,951 & 18,261 & 31,211\\
 & DRIFT & 3,819 & 3,816 & 3,814 & 2,537 & 3,477 & 6,015\\
\midrule
AmbiQT & LoC & 3,046 & 3,046 & 3,044 & 7,499 & 12,199 & 19,698\\
 & DRIFT & 3,046 & 3,044 & 3,044 & 2,699 & 3,304 & 6,003\\
\bottomrule
\end{tabular}
\end{table*}

\subsection{Statistical Comparisons}
\label{app:statistics}
For paired comparisons, each input contributes its score difference between conditions. The AMBROSIA interpretation study resamples 295 database clusters with replacement for 10,000 bootstrap draws, using seed 20261006. Each draw averages the differences over cases in the sampled clusters. Pointwise 95\% percentile intervals cover all three interpretation contrasts and six score metrics. Holm correction covers the three AMBROSIA F1 comparisons. Adjusted p values are 0.00030 for structured interpretation versus AutoCedar, 0.00030 for free interpretation versus AutoCedar, and 0.7555 for structured versus free interpretation. AmbiQT uses a post hoc bootstrap with 50,000 draws over 20 database clusters and seed 20261008. Table~\ref{tab:interpretation-intervals} reports all six metrics for each contrast.

For the direct K3--K2.6 comparison, we pair all inputs with numeric F1 under both reviewers: 3,432 AMBROSIA inputs in 760 database clusters and 2,591 AmbiQT inputs in 20 clusters. AMBROSIA clusters use the SHA-256 of the exact source database dump; AmbiQT uses the source database identifier. We sample the original number of clusters with replacement and retain every paired input in each sampled cluster, including repeated draws. Each replicate divides the summed K3-minus-K2.6 differences by its sampled input count. We use 10,000 draws with seed 2026101212 for AMBROSIA and 50,000 with seed 20261008 for AmbiQT, with NumPy PCG64 and lexicographically ordered clusters. Table~\ref{tab:reviewer-direct-intervals} gives the mean differences and pointwise 95\% percentile intervals, conditional on the saved outputs.

\onecolumn
\section{Detailed Results}
\label{app:diagnostics}
\label{app:fullresults}
Table~\ref{tab:paired} reports matched workflow contrasts in the expanded evaluation. Table~\ref{tab:expanded-complete} collects its conditions and coverage. Table~\ref{tab:strata} separates paired workflow effects by ambiguity group. Table~\ref{tab:complete} retains the initial study and prior interpretation results. One found measures cases with at least one matched target.

\begin{figure}[!ht]
\centering
\includegraphics[width=0.48\textwidth]{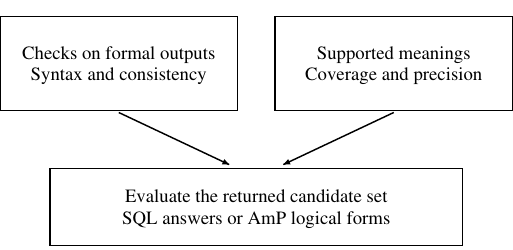}
\caption{The evaluation connects checks on formal outputs with recovery of supported meanings. SQL execution and AmP canonical matching provide the task specific comparisons used for precision and recall.}
\label{fig:venn}
\end{figure}

\begin{table*}[!ht]
\centering\small
\caption{Paired interpretation differences and pointwise 95\% database cluster intervals in percentage points. Base denotes AutoCedar. Each AMBROSIA comparison uses 381 inputs in 295 clusters, and AmbiQT uses 282 inputs in 20 clusters.}
\label{tab:interpretation-intervals}
\begin{tabular}{@{}lllrr@{}}
\toprule
Dataset & Comparison & Metric & Difference & 95\% interval\\
\midrule
AMBROSIA & Structured vs. base & F1 & $-6.44$ & $[-9.07,-3.94]$\\
 &  & Precision & $-7.26$ & $[-9.80,-4.84]$\\
 &  & Recall & $-1.88$ & $[-5.59,+1.82]$\\
 &  & One found & $-2.36$ & $[-6.12,+1.35]$\\
 &  & All found & $-1.31$ & $[-5.32,+2.79]$\\
 &  & Exact set & $-5.51$ & $[-8.56,-2.67]$\\
\midrule
AMBROSIA & Free vs. base & F1 & $-6.81$ & $[-9.23,-4.40]$\\
 &  & Precision & $-9.70$ & $[-12.25,-7.25]$\\
 &  & Recall & $+4.94$ & $[+1.80,+8.20]$\\
 &  & One found & $+4.46$ & $[+1.35,+7.69]$\\
 &  & All found & $+5.51$ & $[+1.61,+9.44]$\\
 &  & Exact set & $-11.02$ & $[-14.48,-7.75]$\\
\midrule
AMBROSIA & Structured vs. free & F1 & $+0.37$ & $[-1.98,+2.65]$\\
 &  & Precision & $+2.44$ & $[+0.07,+4.78]$\\
 &  & Recall & $-6.82$ & $[-9.98,-3.77]$\\
 &  & One found & $-6.82$ & $[-10.22,-3.59]$\\
 &  & All found & $-6.82$ & $[-10.19,-3.49]$\\
 &  & Exact set & $+5.51$ & $[+2.57,+8.44]$\\
\midrule
AmbiQT & Structured vs. base & F1 & $-6.95$ & $[-8.95,-4.97]$\\
 &  & Precision & $-8.84$ & $[-11.38,-6.25]$\\
 &  & Recall & $-2.48$ & $[-4.82,-0.26]$\\
 &  & One found & $-4.26$ & $[-8.30,+0.00]$\\
 &  & All found & $-0.71$ & $[-4.28,+2.59]$\\
 &  & Exact set & $-0.71$ & $[-5.62,+3.73]$\\
\bottomrule
\end{tabular}
\end{table*}

\begin{table*}[!ht]
\centering\small
\caption{AMBROSIA F1 by study and the clear and ambiguous labels supplied by the dataset. Difference is ambiguous F1 minus clear F1, in percentage points. Each row reports its own scored counts.}
\label{tab:interpretation-groups}
\begin{tabular}{@{}llrrrrr@{}}
\toprule
Study & Workflow & Clear $N$ & Clear F1 & Ambiguous $N$ & Ambiguous F1 & Difference\\
\midrule
 Workflows & Direct & 2400 & 58.36 & 1031 & 41.68 & $-16.68$\\
 & AutoCedar & 2400 & 53.61 & 1032 & 42.14 & $-11.46$\\
 & Roundtrip & 2409 & 42.26 & 1024 & 30.16 & $-12.10$\\
 & Clover & 2402 & 43.90 & 1028 & 35.23 & $-8.67$\\
 & IntentPrompt & 2407 & 44.84 & 1024 & 38.51 & $-6.34$\\
 & AutoCedar + Kimi K2.6 & 2400 & 55.18 & 1032 & 42.57 & $-12.61$\\
 & AutoCedar + Kimi K3 & 2400 & 56.07 & 1032 & 42.82 & $-13.26$\\
\midrule
Interpretation & AutoCedar & 266 & 54.51 & 115 & 39.73 & $-14.78$\\
 & AutoCedar + free interpretation & 266 & 46.67 & 115 & 35.31 & $-11.36$\\
 & AutoCedar + structured interpretation & 266 & 48.05 & 115 & 33.33 & $-14.72$\\
\bottomrule
\end{tabular}
\end{table*}
\begin{table}[!ht]
\centering\small
\caption{AmP precision, recall, and canonical set F1 on all 1,000 shared ambiguous inputs. Mean tokens use the 995 shared inputs with complete provider usage across all workflows. Table~\ref{tab:token-coverage} gives the cost accounting.}
\label{tab:amp}
\begin{tabular}{@{}lrrrrr@{}}
\toprule
Workflow & Scored & F1 & Precision & Recall & Mean tokens\\
\midrule
Direct & 1,000 & 73.05 & 70.88 & 76.85 & 7,485\\
IntentPrompt & 1,000 & 73.01 & 69.64 & 78.50 & 8,443\\
AutoCedar & 1,000 & 65.52 & 72.26 & 62.35 & 52,302\\
Clover & 1,000 & 47.12 & 50.45 & 47.20 & 123,860\\
Roundtrip & 1,000 & 73.51 & 70.84 & 78.25 & 13,848\\
LoC & 1,000 & 69.70 & 69.42 & 70.40 & 139,564\\
DRIFT & 1,000 & \textbf{89.02} & 88.09 & 91.55 & 15,199\\
\bottomrule
\end{tabular}
\end{table}

\begin{table*}[t]
\centering\small
\caption{Expanded F1 differences on inputs scored by both methods, in percentage points. Each row uses its own shared input set.}
\label{tab:paired}
\begin{tabular}{@{}llrr@{}}
\toprule
Dataset & Difference vs. Direct & Pairs & $\Delta$F1\\
\midrule
AMBROSIA & AutoCedar & 3,087 & $-3.45$\\
 & Clover & 3,085 & $-11.86$\\
 & IntentPrompt & 3,083 & $-10.96$\\
 & Roundtrip & 3,086 & $-15.06$\\
AmbiQT & AutoCedar & 2,339 & $-6.99$\\
 & Clover & 2,589 & $-40.09$\\
 & IntentPrompt & 2,326 & $-18.64$\\
 & Roundtrip & 2,592 & $-16.27$\\
\bottomrule
\end{tabular}
\end{table*}

\begin{table*}[!ht]
\centering\small
\caption{Expanded AMBROSIA paired F1 changes relative to Direct, in percentage points. Difference is the ambiguous group change minus the clear group change. }
\label{tab:strata}
\begin{tabular}{@{}lrrrrr@{}}
\toprule
Workflow vs. Direct & Clear pairs & $\Delta$F1 clear & Ambiguous pairs & $\Delta$F1 ambiguous & Difference\\
\midrule
AutoCedar + Kimi K2.6 & 2154 & $-3.57$ & 933 & $+0.54$ & $+4.11$\\
AutoCedar + Kimi K3 & 2154 & $-2.69$ & 933 & $+0.83$ & $+3.53$\\
AutoCedar & 2154 & $-5.02$ & 933 & $+0.18$ & $+5.20$\\
Roundtrip & 2167 & $-16.44$ & 919 & $-11.79$ & $+4.65$\\
Clover & 2161 & $-14.14$ & 924 & $-6.52$ & $+7.62$\\
IntentPrompt & 2162 & $-13.96$ & 921 & $-3.91$ & $+10.05$\\
\bottomrule
\end{tabular}
\end{table*}
% BEGIN APPENDIX TABLES
\makeatletter\setlength{\@dblfptop}{0pt}\setlength{\@dblfpsep}{20pt plus 2pt minus 2pt}\makeatother

\begin{table*}[!ht]
\centering\small
\setlength{\tabcolsep}{3pt}
\caption{Initial workflow and confirmation results, together with the prior interpretation study in percent. These runs are separate from the expanded evaluation. AmbiQT interpretation reuses the initial workflow baseline.}
\label{tab:complete}
\begin{tabular}{@{}llrrrrrrrr@{}}
\toprule
Dataset & Workflow & Selected & \textbf{Scored} & F1 & Precision & Recall & One found & All found & Exact set\\
\midrule
\multicolumn{10}{l}{\textit{Workflows}}\\
AMBROSIA & Direct & 382 & 382 & 54.23 & 52.30 & 62.87 & 73.04 & 52.88 & 31.15\\
AMBROSIA & AutoCedar & 382 & 381 & 49.44 & 44.38 & 65.27 & 75.07 & 54.86 & 23.10\\
AMBROSIA & Clover & 382 & 382 & 42.61 & 37.11 & 60.03 & 67.54 & 52.09 & 14.40\\
AMBROSIA & IntentPrompt & 382 & 382 & 39.70 & 31.89 & 61.69 & 69.63 & 53.66 & 9.42\\
AMBROSIA & Roundtrip & 382 & 380 & 34.55 & 31.46 & 43.07 & 47.89 & 38.16 & 18.42\\
AmbiQT & Direct & 305 & 281 & 62.86 & 61.04 & 70.11 & 82.21 & 58.01 & 29.18\\
AmbiQT & AutoCedar & 305 & 282 & 51.67 & 48.33 & 62.23 & 73.76 & 50.71 & 13.83\\
AmbiQT & IntentPrompt & 305 & 289 & 40.45 & 36.01 & 50.69 & 60.21 & 41.18 & 7.61\\
\midrule
\multicolumn{10}{l}{\textit{Interpretation}}\\
AMBROSIA & AutoCedar & 382 & 381 & 50.05 & 44.41 & 67.80 & 76.64 & 58.53 & 23.88\\
AMBROSIA & AutoCedar + free interpretation & 382 & 381 & 43.24 & 34.70 & 72.75 & 81.10 & 64.04 & 12.86\\
AMBROSIA & AutoCedar + structured interpretation & 382 & 381 & 43.61 & 37.15 & 65.92 & 74.28 & 57.22 & 18.37\\
AmbiQT & AutoCedar & 305 & 282 & 51.67 & 48.33 & 62.23 & 73.76 & 50.71 & 13.83\\
AmbiQT & AutoCedar + structured interpretation & 305 & 282 & 44.72 & 39.49 & 59.75 & 69.50 & 50.00 & 13.12\\
\midrule
\multicolumn{10}{l}{\textit{Confirmation}}\\
AMBROSIA & No extra model & 382 & 381 & 49.03 & 43.90 & 64.83 & 74.54 & 54.33 & 21.78\\
AMBROSIA & Kimi K3 & 382 & 381 & 50.62 & 46.09 & 64.87 & 75.07 & 54.33 & 24.93\\
AMBROSIA & Kimi K2.6 & 382 & 380 & 49.54 & 44.88 & 64.47 & 73.68 & 55.00 & 23.42\\
AmbiQT & No extra model & 305 & 282 & 51.36 & 48.81 & 60.99 & 73.76 & 48.23 & 14.54\\
AmbiQT & Kimi K3 & 305 & 282 & 52.86 & 51.21 & 60.64 & 72.34 & 48.94 & 18.44\\
AmbiQT & Kimi K2.6 & 305 & 281 & 51.32 & 48.97 & 60.32 & 72.24 & 48.40 & 16.37\\
\bottomrule
\end{tabular}
\end{table*}

\begin{table*}[!ht]
\centering\small
\setlength{\tabcolsep}{3pt}
\caption{Expanded coverage and answer set recovery in percent. Scored counts eligible records with available scores. Ineligible references and other unscored cases complete the selected denominator. Main paper Table~\ref{tab:main} and Supplementary Table~\ref{tab:confirmation} give F1, precision, and recall for these same conditions.}
\label{tab:expanded-complete}
\begin{tabular}{@{}llrrrrrrr@{}}
\toprule
Dataset & Workflow & Selected & \textbf{Scored} & Ineligible & Other unscored & One found & All found & Exact set\\
\midrule
AMBROSIA & Direct & 3,437 & 3,431 & 6 & 0 & 71.70 & 52.78 & 29.41\\
 & AutoCedar & 3,437 & 3,432 & 5 & 0 & 76.63 & 58.54 & 22.79\\
 & Clover & 3,437 & 3,430 & 6 & 1 & 65.04 & 50.35 & 14.75\\
 & IntentPrompt & 3,437 & 3,431 & 6 & 0 & 70.94 & 53.63 & 15.10\\
 & Roundtrip & 3,437 & 3,433 & 4 & 0 & 53.51 & 41.30 & 20.07\\
 & AutoCedar + Kimi K3 & 3,437 & 3,432 & 5 & 0 & 76.78 & 57.69 & 25.23\\
 & AutoCedar + Kimi K2.6 & 3,437 & 3,432 & 5 & 0 & 76.78 & 58.39 & 23.95\\
\midrule
AmbiQT & Direct & 2,741 & 2,592 & 149 & 0 & 79.75 & 57.60 & 27.66\\
 & AutoCedar & 2,741 & 2,591 & 150 & 0 & 76.69 & 54.69 & 17.02\\
 & Clover & 3,046 & 2,871 & 170 & 5 & 35.14 & 17.97 & 2.86\\
 & IntentPrompt & 2,741 & 2,585 & 156 & 0 & 62.63 & 43.17 & 8.32\\
 & Roundtrip & 3,046 & 2,874 & 172 & 0 & 58.84 & 45.79 & 20.63\\
 & AutoCedar + Kimi K3 & 2,741 & 2,591 & 150 & 0 & 76.42 & 54.26 & 20.65\\
 & AutoCedar + Kimi K2.6 & 2,741 & 2,591 & 150 & 0 & 75.96 & 53.26 & 18.14\\
\bottomrule
\end{tabular}
\end{table*}

\begin{table*}[!ht]
\centering\small
\caption{Five SQL workflows evaluated on the same scored inputs within each dataset. Values are percentages computed from saved scores. Bold marks the highest value in each dataset.}
\label{tab:common-five}
\begin{tabular}{@{}llrrrr@{}}
\toprule
Dataset & Workflow & Shared inputs & F1 & Precision & Recall\\
\midrule
AMBROSIA & Direct & 2,249 & \textbf{53.72} & \textbf{51.63} & 62.35\\
 & AutoCedar & 2,249 & 50.17 & 45.27 & \textbf{67.27}\\
 & Roundtrip & 2,249 & 38.20 & 35.29 & 46.47\\
 & Clover & 2,249 & 41.58 & 35.94 & 58.40\\
 & IntentPrompt & 2,249 & 42.80 & 36.19 & 62.01\\
\midrule
AmbiQT & Direct & 2,108 & \textbf{61.33} & \textbf{59.30} & \textbf{69.21}\\
 & AutoCedar & 2,108 & 54.34 & 50.72 & 65.65\\
 & Roundtrip & 2,108 & 44.63 & 41.60 & 52.16\\
 & Clover & 2,108 & 20.48 & 18.99 & 26.28\\
 & IntentPrompt & 2,108 & 42.30 & 38.00 & 52.87\\
\bottomrule
\end{tabular}
\end{table*}

The five workflows share 2,254 selected AMBROSIA IDs and 2,226 selected AmbiQT IDs before score filtering. Requiring a valid score from every workflow leaves 2,249 and 2,108 inputs respectively. Table~\ref{tab:common-five} uses these fixed intersections, while Table~\ref{tab:paired} uses a separate intersection for each pair.

Table~\ref{tab:confirmation} reports the condition means underlying the paired feedback contrasts in Main paper Table~\ref{tab:confirmation-paired}.

\begin{table}[!ht]
\centering\small
\setlength{\tabcolsep}{4pt}
\caption{Review feedback results in percent. P and R denote precision and recall. Each mean uses the scored inputs for that condition. Bold marks the highest mean for each dataset.}
\label{tab:confirmation}
\begin{tabular}{@{}lrrrrr@{}}
\toprule
Reviewer & Scored & F1 & P & R & Exact set\\
\midrule
\multicolumn{6}{l}{AMBROSIA}\\
None & 3,432 & 50.16 & 45.06 & \textbf{67.81} & 22.79\\
Kimi K3 & 3,432 & \textbf{52.09} & \textbf{47.91} & 67.36 & \textbf{25.23}\\
Kimi K2.6 & 3,432 & 51.39 & 46.72 & 67.68 & 23.95\\
\midrule
\multicolumn{6}{l}{AmbiQT}\\
None & 2,591 & 54.45 & 50.89 & \textbf{65.69} & 17.02\\
Kimi K3 & 2,591 & \textbf{56.49} & \textbf{54.42} & 65.34 & \textbf{20.65}\\
Kimi K2.6 & 2,591 & 54.69 & 52.02 & 64.61 & 18.14\\
\bottomrule
\end{tabular}
\end{table}

\begin{table}[!ht]
\centering\small
\setlength{\tabcolsep}{4pt}
\caption{Direct K3 minus K2.6 paired F1 differences, in percentage points. Intervals resample database clusters as described in Section~\ref{app:statistics}.}
\label{tab:reviewer-direct-intervals}
\begin{tabular}{@{}lrrrr@{}}
\toprule
Dataset & Pairs & Clusters & $\Delta$F1 & 95\% interval\\
\midrule
AMBROSIA & 3,432 & 760 & $+0.70$ & $[0.31,1.09]$\\
AmbiQT & 2,591 & 20 & $+1.80$ & $[1.25,2.46]$\\
\bottomrule
\end{tabular}
\end{table}

\clearpage
\subsection{Interfaces and Supplementary Visual Summaries}
\label{app:visual-summaries}
\begin{figure}[!ht]
\centering
\includegraphics[width=0.58\linewidth]{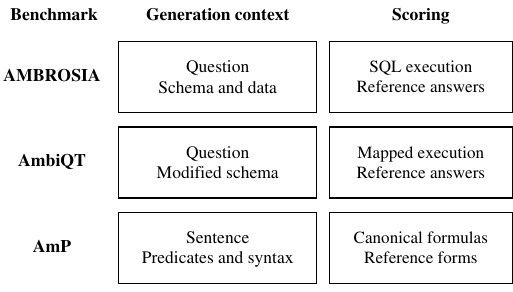}
\caption{Inputs and scoring for all three benchmarks. Reference outputs enter at evaluation.}
\label{fig:io}
\end{figure}

\begin{figure}[!ht]
\centering
\includegraphics[width=0.48\linewidth]{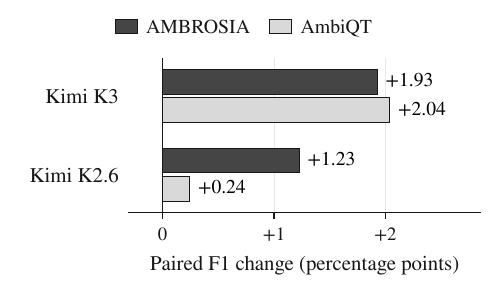}
\caption{Paired F1 gains from review feedback relative to AutoCedar without an extra model. Values are percentage points on the shared inputs in Main paper Table~\ref{tab:confirmation-paired}.}
\label{fig:bar}
\end{figure}

Figure~\ref{fig:io} summarizes the benchmark interfaces. Figure~\ref{fig:bar} visualizes the paired review gains in Main paper Table~\ref{tab:confirmation-paired}; Main paper Table~\ref{tab:layers} reports the interpretation cost comparison.

\clearpage
\begin{table*}[p]

\section{Selected Examples}

\label{app:samples}

Each comparison contains ten fixed examples. The expanded workflow examples include all seven generation and confirmation conditions on five AMBROSIA and five AmbiQT inputs. The interpretation and initial confirmation comparisons each include ten examples.

Each table gives the question, reference SQL, precision, recall, F1, and the first returned SQL candidate. Scores use the complete answer set, with the total candidate count shown beside the example. Intermediate entries show saved readings, structured descriptions, Boolean atoms, or reconstruction outputs. 
\medskip

\centering\footnotesize

\setlength{\tabcolsep}{3pt}

\caption{Workflow comparison. Clear input. Precision, recall, and F1 are percentages computed from the complete returned set against the benchmark references.}

\label{tab:sample-evidence-1}
\begin{tabular}{@{}p{0.19\textwidth}rrrp{0.57\textwidth}@{}}

\toprule

\multicolumn{5}{@{}p{0.98\textwidth}@{}}{\textbf{AMBROSIA 426}. What are the chefs who speak English and servers who speak English.}\\[4pt]

\multicolumn{5}{@{}p{0.98\textwidth}@{}}{\textit{Reference SQL.} {\ttfamily\raggedright SELECT Chefs.\allowbreak{}lastname FROM Languages JOIN Chefs ON Languages.\allowbreak{}LanguageID = Chefs.\allowbreak{}LanguageID WHERE Languages.\allowbreak{}LanguageName = "English" UNION SELECT Servers.\allowbreak{}lastname FROM Languages JOIN Servers ON Languages.\allowbreak{}LanguageID = Servers.\allowbreak{}LanguageID WHERE Languages.\allowbreak{}LanguageName = "English"}}\\[5pt]

Workflow & P & R & F1 & First returned SQL and candidate count\\

\midrule

Direct & 0.00 & 0.00 & 0.00 & {\ttfamily\raggedright SELECT c.\allowbreak{}FirstName,\allowbreak{} c.\allowbreak{}LastName FROM Chefs c JOIN Languages l ON c.\allowbreak{}LanguageID = l.\allowbreak{}LanguageID WHERE l.\allowbreak{}LanguageName = 'English' UNION SELECT s.\allowbreak{}FirstName,\allowbreak{} s.\allowbreak{}LastName FROM Servers s JOIN Languages l ON s.\allowbreak{}LanguageID = l.\allowbreak{}LanguageID WHERE l.\allowbreak{}LanguageName = 'English'}, 1 candidate\\[4pt]

AutoCedar & 0.00 & 0.00 & 0.00 & {\ttfamily\raggedright SELECT FirstName,\allowbreak{} LastName FROM Chefs JOIN Languages ON Chefs.\allowbreak{}LanguageID = Languages.\allowbreak{}LanguageID WHERE Languages.\allowbreak{}LanguageName = 'English' UNION SELECT FirstName,\allowbreak{} LastName FROM Servers JOIN Languages ON Servers.\allowbreak{}LanguageID = Languages.\allowbreak{}LanguageID WHERE Languages.\allowbreak{}LanguageName = 'English'}, 1 candidate\\[4pt]

Roundtrip & 0.00 & 0.00 & 0.00 & Same first query as Direct, 4 candidates\\[4pt]

Clover & 0.00 & 0.00 & 0.00 & Empty returned set, 0 candidates\\[4pt]

IntentPrompt & 0.00 & 0.00 & 0.00 & {\ttfamily\raggedright SELECT 'Chef' AS Role,\allowbreak{} c.\allowbreak{}FirstName,\allowbreak{} c.\allowbreak{}LastName FROM Chefs c JOIN Languages l ON c.\allowbreak{}LanguageID = l.\allowbreak{}LanguageID WHERE l.\allowbreak{}LanguageName = 'English' UNION ALL SELECT 'Server' AS Role,\allowbreak{} s.\allowbreak{}FirstName,\allowbreak{} s.\allowbreak{}LastName FROM Servers s JOIN Languages l ON s.\allowbreak{}LanguageID = l.\allowbreak{}LanguageID WHERE l.\allowbreak{}LanguageName = 'English'}, 3 candidates\\[4pt]

AutoCedar + Kimi K3 & 0.00 & 0.00 & 0.00 & Same first query as AutoCedar, 1 candidate\\[4pt]

AutoCedar + Kimi K2.6 & 0.00 & 0.00 & 0.00 & Same first query as AutoCedar, 1 candidate\\[4pt]

\midrule

\multicolumn{5}{l}{\textit{Recorded intermediate representations}}\\[3pt]

\multicolumn{5}{@{}p{0.98\textwidth}@{}}{\textbf{Roundtrip.} Reconstructed reading Return the FirstName and LastName of every chef whose LanguageID matches a Languages row with LanguageName 'English', together with the FirstName and LastName of every server whose LanguageID matches a Languages row (excerpt)}\\[4pt]

\multicolumn{5}{@{}p{0.98\textwidth}@{}}{\textbf{Clover.} All four proposed branches failed a description-to-specification or description-to-code check, so the returned set was empty.}\\[4pt]

\bottomrule

\end{tabular}

\end{table*}

\begin{table*}[p]

\centering\footnotesize

\setlength{\tabcolsep}{3pt}

\caption{Workflow comparison. Clear input. Precision, recall, and F1 are percentages computed from the complete returned set against the benchmark references.}

\label{tab:sample-evidence-2}
\begin{tabular}{@{}p{0.19\textwidth}rrrp{0.57\textwidth}@{}}

\toprule

\multicolumn{5}{@{}p{0.98\textwidth}@{}}{\textbf{AMBROSIA 428}. Can you show me the sizes of all hot tubs in Miami along with the sizes of all swimming pools in any given location.}\\[4pt]

\multicolumn{5}{@{}p{0.98\textwidth}@{}}{\textit{Reference SQL.} {\ttfamily\raggedright SELECT SwimmingPool.\allowbreak{}Size FROM City JOIN Hotel ON City.\allowbreak{}ID = Hotel.\allowbreak{}ID JOIN SwimmingPool ON Hotel.\allowbreak{}ID = SwimmingPool.\allowbreak{}ID UNION SELECT HotTub.\allowbreak{}Capacity FROM City JOIN Hotel ON City.\allowbreak{}ID = Hotel.\allowbreak{}ID JOIN HotTub ON Hotel.\allowbreak{}ID = HotTub.\allowbreak{}ID WHERE City.\allowbreak{}Name = 'Miami'}}\\[5pt]

Workflow & P & R & F1 & First returned SQL and candidate count\\

\midrule

Direct & 33.33 & 100.00 & 50.00 & {\ttfamily\raggedright SELECT HotTub.\allowbreak{}Capacity AS Size FROM HotTub JOIN Hotel ON HotTub.\allowbreak{}ID = Hotel.\allowbreak{}ID JOIN City ON Hotel.\allowbreak{}ID = City.\allowbreak{}ID WHERE City.\allowbreak{}Name = 'Miami' UNION ALL SELECT SwimmingPool.\allowbreak{}Size AS Size FROM SwimmingPool}, 3 candidates\\[4pt]

AutoCedar & 50.00 & 100.00 & 66.67 & {\ttfamily\raggedright SELECT h.\allowbreak{}Capacity AS Size FROM HotTub h JOIN Hotel ho ON h.\allowbreak{}ID = ho.\allowbreak{}ID JOIN City c ON ho.\allowbreak{}ID = c.\allowbreak{}ID WHERE c.\allowbreak{}Name = 'Miami' UNION ALL SELECT s.\allowbreak{}Size AS Size FROM SwimmingPool s}, 5 candidates\\[4pt]

Roundtrip & 0.00 & 0.00 & 0.00 & {\ttfamily\raggedright SELECT 'Hot Tub' AS Type,\allowbreak{} ht.\allowbreak{}Capacity AS Size,\allowbreak{} c.\allowbreak{}Name AS Location FROM HotTub ht JOIN Hotel h ON ht.\allowbreak{}ID = h.\allowbreak{}ID JOIN City c ON h.\allowbreak{}ID = c.\allowbreak{}ID WHERE c.\allowbreak{}Name = 'Miami' UNION ALL SELECT 'Swimming Pool' AS Type,\allowbreak{} sp.\allowbreak{}Size AS Size,\allowbreak{} c.\allowbreak{}Name AS Location FROM SwimmingPool sp JOIN Hotel h ON sp.\allowbreak{}ID = h.\allowbreak{}ID JOIN City c ON h.\allowbreak{}ID = c.\allowbreak{}ID}, 2 candidates\\[4pt]

Clover & 0.00 & 0.00 & 0.00 & {\ttfamily\raggedright SELECT HotTub.\allowbreak{}Capacity AS Size FROM HotTub JOIN Hotel ON HotTub.\allowbreak{}ID = Hotel.\allowbreak{}ID JOIN City ON Hotel.\allowbreak{}ID = City.\allowbreak{}ID WHERE City.\allowbreak{}Name = 'Miami' UNION ALL SELECT SwimmingPool.\allowbreak{}Size AS Size FROM SwimmingPool JOIN Hotel ON SwimmingPool.\allowbreak{}ID = Hotel.\allowbreak{}ID JOIN City ON Hotel.\allowbreak{}ID = City.\allowbreak{}ID WHERE City.\allowbreak{}Name = 'Miami'}, 3 candidates\\[4pt]

IntentPrompt & 20.00 & 100.00 & 33.33 & {\ttfamily\raggedright SELECT 'Hot Tub' AS Type,\allowbreak{} City.\allowbreak{}Name AS Location,\allowbreak{} HotTub.\allowbreak{}Capacity AS Size FROM HotTub JOIN Hotel ON HotTub.\allowbreak{}ID = Hotel.\allowbreak{}ID JOIN City ON Hotel.\allowbreak{}ID = City.\allowbreak{}ID WHERE City.\allowbreak{}Name = 'Miami' UNION ALL SELECT 'Swimming Pool' AS Type,\allowbreak{} City.\allowbreak{}Name AS Location,\allowbreak{} SwimmingPool.\allowbreak{}Size AS Size FROM SwimmingPool JOIN Hotel ON SwimmingPool.\allowbreak{}ID = Hotel.\allowbreak{}ID JOIN City ON Hotel.\allowbreak{}ID = City.\allowbreak{}ID}, 5 candidates\\[4pt]

AutoCedar + Kimi K3 & 50.00 & 100.00 & 66.67 & Same first query as AutoCedar, 5 candidates\\[4pt]

AutoCedar + Kimi K2.6 & 50.00 & 100.00 & 66.67 & Same first query as AutoCedar, 5 candidates\\[4pt]

\midrule

\multicolumn{5}{l}{\textit{Recorded intermediate representations}}\\[3pt]

\multicolumn{5}{@{}p{0.98\textwidth}@{}}{\textbf{Roundtrip.} Reconstructed reading This query returns a combined result set with three columns, Type, Size, and Location. For the first part, it selects one row for every HotTub whose joined Hotel and City match on ID and whose City.Name is exactly (excerpt)}\\[4pt]

\multicolumn{5}{@{}p{0.98\textwidth}@{}}{\textbf{Clover.} SQL reconstruction and specification checks mismatched. The reconstructed description was assessed by the model.}\\[4pt]

\bottomrule

\end{tabular}

\end{table*}

\begin{table*}[p]

\centering\footnotesize

\setlength{\tabcolsep}{3pt}

\caption{Workflow comparison. Ambiguous input. Precision, recall, and F1 are percentages computed from the complete returned set against the benchmark references.}

\label{tab:sample-evidence-3}
\begin{tabular}{@{}p{0.19\textwidth}rrrp{0.57\textwidth}@{}}

\toprule

\multicolumn{5}{@{}p{0.98\textwidth}@{}}{\textbf{AMBROSIA 429}. What size are all swimming pools and hot tubs in Miami.}\\[4pt]

\multicolumn{5}{@{}p{0.98\textwidth}@{}}{\textit{Reference SQL 1.} {\ttfamily\raggedright SELECT SwimmingPool.\allowbreak{}Size FROM City JOIN Hotel ON City.\allowbreak{}ID = Hotel.\allowbreak{}ID JOIN SwimmingPool ON Hotel.\allowbreak{}ID = SwimmingPool.\allowbreak{}ID WHERE City.\allowbreak{}Name = 'Miami' UNION SELECT HotTub.\allowbreak{}Capacity FROM City JOIN Hotel ON City.\allowbreak{}ID = Hotel.\allowbreak{}ID JOIN HotTub ON Hotel.\allowbreak{}ID = HotTub.\allowbreak{}ID WHERE City.\allowbreak{}Name = 'Miami'}\par\smallskip \textit{Reference SQL 2.} {\ttfamily\raggedright SELECT SwimmingPool.\allowbreak{}Size FROM City JOIN Hotel ON City.\allowbreak{}ID = Hotel.\allowbreak{}ID JOIN SwimmingPool ON Hotel.\allowbreak{}ID = SwimmingPool.\allowbreak{}ID UNION SELECT HotTub.\allowbreak{}Capacity FROM City JOIN Hotel ON City.\allowbreak{}ID = Hotel.\allowbreak{}ID JOIN HotTub ON Hotel.\allowbreak{}ID = HotTub.\allowbreak{}ID WHERE City.\allowbreak{}Name = 'Miami'}}\\[5pt]

Workflow & P & R & F1 & First returned SQL and candidate count\\

\midrule

Direct & 100.00 & 50.00 & 66.67 & {\ttfamily\raggedright SELECT sp.\allowbreak{}Size FROM SwimmingPool sp JOIN Hotel h ON sp.\allowbreak{}ID = h.\allowbreak{}ID JOIN City c ON h.\allowbreak{}ID = c.\allowbreak{}ID WHERE c.\allowbreak{}Name = 'Miami' UNION ALL SELECT ht.\allowbreak{}Capacity FROM HotTub ht JOIN Hotel h ON ht.\allowbreak{}ID = h.\allowbreak{}ID JOIN City c ON h.\allowbreak{}ID = c.\allowbreak{}ID WHERE c.\allowbreak{}Name = 'Miami'}, 1 candidate\\[4pt]

AutoCedar & 50.00 & 100.00 & 66.67 & {\ttfamily\raggedright SELECT SwimmingPool.\allowbreak{}Size FROM SwimmingPool JOIN Hotel ON SwimmingPool.\allowbreak{}ID = Hotel.\allowbreak{}ID JOIN City ON Hotel.\allowbreak{}ID = City.\allowbreak{}ID WHERE City.\allowbreak{}Name = 'Miami' UNION ALL SELECT HotTub.\allowbreak{}Capacity FROM HotTub JOIN Hotel ON HotTub.\allowbreak{}ID = Hotel.\allowbreak{}ID JOIN City ON Hotel.\allowbreak{}ID = City.\allowbreak{}ID WHERE City.\allowbreak{}Name = 'Miami'}, 5 candidates\\[4pt]

Roundtrip & 0.00 & 0.00 & 0.00 & {\ttfamily\raggedright SELECT 'Swimming Pool' AS FacilityType,\allowbreak{} sp.\allowbreak{}Size AS Size FROM City c JOIN Hotel h ON h.\allowbreak{}ID = c.\allowbreak{}ID JOIN SwimmingPool sp ON sp.\allowbreak{}ID = h.\allowbreak{}ID WHERE c.\allowbreak{}Name = 'Miami' UNION ALL SELECT 'Hot Tub' AS FacilityType,\allowbreak{} CAST(ht.\allowbreak{}Capacity AS TEXT) AS Size FROM City c JOIN Hotel h ON h.\allowbreak{}ID = c.\allowbreak{}ID JOIN HotTub ht ON ht.\allowbreak{}ID = h.\allowbreak{}ID WHERE c.\allowbreak{}Name = 'Miami'}, 3 candidates\\[4pt]

Clover & 25.00 & 50.00 & 33.33 & {\ttfamily\raggedright SELECT sp.\allowbreak{}Size AS Size FROM SwimmingPool sp JOIN Hotel h ON sp.\allowbreak{}ID = h.\allowbreak{}ID JOIN City c ON h.\allowbreak{}ID = c.\allowbreak{}ID WHERE c.\allowbreak{}Name = 'Miami' UNION ALL SELECT NULL AS Size FROM HotTub ht JOIN Hotel h ON ht.\allowbreak{}ID = h.\allowbreak{}ID JOIN City c ON h.\allowbreak{}ID = c.\allowbreak{}ID WHERE c.\allowbreak{}Name = 'Miami'}, 4 candidates\\[4pt]

IntentPrompt & 0.00 & 0.00 & 0.00 & {\ttfamily\raggedright SELECT 'Swimming Pool' AS Type,\allowbreak{} sp.\allowbreak{}Size AS Size FROM SwimmingPool sp JOIN Hotel h ON sp.\allowbreak{}ID = h.\allowbreak{}ID JOIN City c ON h.\allowbreak{}ID = c.\allowbreak{}ID WHERE c.\allowbreak{}Name = 'Miami' UNION ALL SELECT 'Hot Tub' AS Type,\allowbreak{} CAST(ht.\allowbreak{}Capacity AS TEXT) AS Size FROM HotTub ht JOIN Hotel h ON ht.\allowbreak{}ID = h.\allowbreak{}ID JOIN City c ON h.\allowbreak{}ID = c.\allowbreak{}ID WHERE c.\allowbreak{}Name = 'Miami'}, 3 candidates\\[4pt]

AutoCedar + Kimi K3 & 33.33 & 50.00 & 40.00 & Same first query as AutoCedar, 5 candidates\\[4pt]

AutoCedar + Kimi K2.6 & 50.00 & 100.00 & 66.67 & Same first query as AutoCedar, 4 candidates\\[4pt]

\midrule

\multicolumn{5}{l}{\textit{Recorded intermediate representations}}\\[3pt]

\multicolumn{5}{@{}p{0.98\textwidth}@{}}{\textbf{Roundtrip.} Reconstructed reading Returns a combined result set using UNION ALL. For every City row named 'Miami' joined to a Hotel row whose ID equals that City row's ID, and joined to a SwimmingPool row whose ID equals that Hotel row's ID, it (excerpt)}\\[4pt]

\multicolumn{5}{@{}p{0.98\textwidth}@{}}{\textbf{Clover.} SQL reconstruction and specification checks matched. The reconstructed description was assessed by the model.}\\[4pt]

\bottomrule

\end{tabular}

\end{table*}

\begin{table*}[p]

\centering\footnotesize

\setlength{\tabcolsep}{3pt}

\caption{Workflow comparison. Clear input. Precision, recall, and F1 are percentages computed from the complete returned set against the benchmark references.}

\label{tab:sample-evidence-4}
\begin{tabular}{@{}p{0.19\textwidth}rrrp{0.57\textwidth}@{}}

\toprule

\multicolumn{5}{@{}p{0.98\textwidth}@{}}{\textbf{AMBROSIA 430}. Provide a list of all family suites and also provide a list of single rooms that cost 150.}\\[4pt]

\multicolumn{5}{@{}p{0.98\textwidth}@{}}{\textit{Reference SQL.} {\ttfamily\raggedright SELECT FamilySuites.\allowbreak{}Room FROM FamilySuites UNION SELECT SingleRooms.\allowbreak{}Room FROM SingleRooms WHERE SingleRooms.\allowbreak{}Price\_\allowbreak{}per\_\allowbreak{}night = 150.\allowbreak{}0}}\\[5pt]

Workflow & P & R & F1 & First returned SQL and candidate count\\

\midrule

Direct & 0.00 & 0.00 & 0.00 & {\ttfamily\raggedright SELECT Room,\allowbreak{} Price\_\allowbreak{}per\_\allowbreak{}night FROM FamilySuites UNION ALL SELECT Room,\allowbreak{} Price\_\allowbreak{}per\_\allowbreak{}night FROM SingleRooms WHERE Price\_\allowbreak{}per\_\allowbreak{}night = 150}, 2 candidates\\[4pt]

AutoCedar & 25.00 & 100.00 & 40.00 & {\ttfamily\raggedright SELECT SuiteID,\allowbreak{} Room,\allowbreak{} OccupancyLimit,\allowbreak{} Amenities,\allowbreak{} Price\_\allowbreak{}per\_\allowbreak{}night,\allowbreak{} NULL AS RoomID,\allowbreak{} NULL AS NumberOfBeds,\allowbreak{} NULL AS View,\allowbreak{} NULL AS Services FROM FamilySuites UNION ALL SELECT NULL,\allowbreak{} Room,\allowbreak{} NULL,\allowbreak{} NULL,\allowbreak{} Price\_\allowbreak{}per\_\allowbreak{}night,\allowbreak{} RoomID,\allowbreak{} NumberOfBeds,\allowbreak{} View,\allowbreak{} Services FROM SingleRooms WHERE Price\_\allowbreak{}per\_\allowbreak{}night = 150}, 7 candidates\\[4pt]

Roundtrip & 0.00 & 0.00 & 0.00 & {\ttfamily\raggedright SELECT Room,\allowbreak{} 'Family Suite' AS Type,\allowbreak{} OccupancyLimit AS Capacity,\allowbreak{} Amenities AS Details,\allowbreak{} Price\_\allowbreak{}per\_\allowbreak{}night FROM FamilySuites UNION ALL SELECT Room,\allowbreak{} 'Single Room' AS Type,\allowbreak{} NumberOfBeds AS Capacity,\allowbreak{} Services AS Details,\allowbreak{} Price\_\allowbreak{}per\_\allowbreak{}night FROM SingleRooms WHERE Price\_\allowbreak{}per\_\allowbreak{}night = 150.\allowbreak{}0}, 4 candidates\\[4pt]

Clover & 33.33 & 100.00 & 50.00 & Same first query as Direct, 3 candidates\\[4pt]

IntentPrompt & 0.00 & 0.00 & 0.00 & {\ttfamily\raggedright SELECT 'Family Suite' AS Type,\allowbreak{} Room,\allowbreak{} Price\_\allowbreak{}per\_\allowbreak{}night FROM FamilySuites UNION ALL SELECT 'Single Room' AS Type,\allowbreak{} Room,\allowbreak{} Price\_\allowbreak{}per\_\allowbreak{}night FROM SingleRooms WHERE Price\_\allowbreak{}per\_\allowbreak{}night = 150 ORDER BY Type,\allowbreak{} Room}, 3 candidates\\[4pt]

AutoCedar + Kimi K3 & 25.00 & 100.00 & 40.00 & Same first query as AutoCedar, 6 candidates\\[4pt]

AutoCedar + Kimi K2.6 & 25.00 & 100.00 & 40.00 & Same first query as AutoCedar, 7 candidates\\[4pt]

\midrule

\multicolumn{5}{l}{\textit{Recorded intermediate representations}}\\[3pt]

\multicolumn{5}{@{}p{0.98\textwidth}@{}}{\textbf{Roundtrip.} Reconstructed reading For every row in FamilySuites, return its Room, the literal 'Family Suite' as Type, its OccupancyLimit as Capacity, its Amenities as Details, and its Price\_per\_night. Also return, for every row in SingleRooms whose (excerpt)}\\[4pt]

\multicolumn{5}{@{}p{0.98\textwidth}@{}}{\textbf{Clover.} SQL reconstruction and specification checks mismatched. The reconstructed description was assessed by the model.}\\[4pt]

\bottomrule

\end{tabular}

\end{table*}

\begin{table*}[p]

\centering\footnotesize

\setlength{\tabcolsep}{3pt}

\caption{Workflow comparison. Clear input. Precision, recall, and F1 are percentages computed from the complete returned set against the benchmark references.}

\label{tab:sample-evidence-5}
\begin{tabular}{@{}p{0.19\textwidth}rrrp{0.57\textwidth}@{}}

\toprule

\multicolumn{5}{@{}p{0.98\textwidth}@{}}{\textbf{AMBROSIA 434}. Can you show chalets and cabins that cost \$200 per night.}\\[4pt]

\multicolumn{5}{@{}p{0.98\textwidth}@{}}{\textit{Reference SQL.} {\ttfamily\raggedright SELECT Chalets.\allowbreak{}chaletName as name FROM Chalets WHERE Chalets.\allowbreak{}pricePerNight = 200.\allowbreak{}0 UNION SELECT Cabins.\allowbreak{}cabinName as name FROM Cabins WHERE Cabins.\allowbreak{}pricePerNight = 200.\allowbreak{}0}}\\[5pt]

Workflow & P & R & F1 & First returned SQL and candidate count\\

\midrule

Direct & 0.00 & 0.00 & 0.00 & {\ttfamily\raggedright SELECT cabinName AS name,\allowbreak{} pricePerNight FROM Cabins WHERE pricePerNight = 200 UNION SELECT chaletName AS name,\allowbreak{} pricePerNight FROM Chalets WHERE pricePerNight = 200}, 3 candidates\\[4pt]

AutoCedar & 20.00 & 100.00 & 33.33 & {\ttfamily\raggedright SELECT chaletName AS name FROM Chalets WHERE pricePerNight = 200 UNION ALL SELECT cabinName AS name FROM Cabins WHERE pricePerNight = 200}, 6 candidates\\[4pt]

Roundtrip & 0.00 & 0.00 & 0.00 & {\ttfamily\raggedright SELECT 'Chalet' AS type,\allowbreak{} id,\allowbreak{} chaletName AS name,\allowbreak{} pricePerNight,\allowbreak{} location,\allowbreak{} capacity FROM Chalets WHERE pricePerNight = 200 UNION ALL SELECT 'Cabin' AS type,\allowbreak{} id,\allowbreak{} cabinName AS name,\allowbreak{} pricePerNight,\allowbreak{} location,\allowbreak{} capacity FROM Cabins WHERE pricePerNight = 200}, 1 candidate\\[4pt]

Clover & 50.00 & 100.00 & 66.67 & {\ttfamily\raggedright SELECT chaletName AS name FROM Chalets WHERE pricePerNight = 200 UNION SELECT cabinName AS name FROM Cabins WHERE pricePerNight = 200}, 2 candidates\\[4pt]

IntentPrompt & 0.00 & 0.00 & 0.00 & {\ttfamily\raggedright SELECT 'Cabin' AS type,\allowbreak{} cabinName AS name FROM Cabins WHERE pricePerNight = 200 UNION ALL SELECT 'Chalet' AS type,\allowbreak{} chaletName AS name FROM Chalets WHERE pricePerNight = 200}, 3 candidates\\[4pt]

AutoCedar + Kimi K3 & 16.67 & 100.00 & 28.57 & Same first query as Roundtrip, 7 candidates\\[4pt]

AutoCedar + Kimi K2.6 & 25.00 & 100.00 & 40.00 & Same first query as AutoCedar, 7 candidates\\[4pt]

\midrule

\multicolumn{5}{l}{\textit{Recorded intermediate representations}}\\[3pt]

\multicolumn{5}{@{}p{0.98\textwidth}@{}}{\textbf{Roundtrip.} Reconstructed reading This query returns a combined result set with columns type, id, name, pricePerNight, location, and capacity. It has two parts joined by UNION ALL. The first part selects every row from the Chalets table whose (excerpt)}\\[4pt]

\multicolumn{5}{@{}p{0.98\textwidth}@{}}{\textbf{Clover.} SQL reconstruction and specification checks mismatched. The reconstructed description was assessed by the model.}\\[4pt]

\bottomrule

\end{tabular}

\end{table*}

\begin{table*}[p]

\centering\footnotesize

\setlength{\tabcolsep}{3pt}

\caption{Workflow comparison. Ambiguous input. Precision, recall, and F1 are percentages computed from the complete returned set against the benchmark references.}

\label{tab:sample-evidence-6}
\begin{tabular}{@{}p{0.19\textwidth}rrrp{0.57\textwidth}@{}}

\toprule

\multicolumn{5}{@{}p{0.98\textwidth}@{}}{\textbf{AMBIQT 000000}. Show name, country, age for all singers ordered by age from the oldest to the youngest.}\\[4pt]

\multicolumn{5}{@{}p{0.98\textwidth}@{}}{\textit{Reference SQL 1.} {\ttfamily\raggedright select artist\_\allowbreak{}name,\allowbreak{} country,\allowbreak{} age from singer order by age desc}\par\smallskip \textit{Reference SQL 2.} {\ttfamily\raggedright select performer\_\allowbreak{}name,\allowbreak{} country,\allowbreak{} age from singer order by age desc}}\\[5pt]

Workflow & P & R & F1 & First returned SQL and candidate count\\

\midrule

Direct & 100.00 & 100.00 & 100.00 & {\ttfamily\raggedright select artist\_\allowbreak{}name,\allowbreak{} country,\allowbreak{} age from singer order by age desc}, 5 candidates\\[4pt]

AutoCedar & 100.00 & 100.00 & 100.00 & Same first query as Direct, 2 candidates\\[4pt]

Roundtrip & 100.00 & 100.00 & 100.00 & {\ttfamily\raggedright SELECT artist\_\allowbreak{}name AS name,\allowbreak{} country,\allowbreak{} age FROM singer ORDER BY age DESC}, 2 candidates\\[4pt]

Clover & 100.00 & 50.00 & 66.67 & Same first query as Roundtrip, 1 candidate\\[4pt]

IntentPrompt & 66.67 & 100.00 & 80.00 & Same first query as Direct, 3 candidates\\[4pt]

AutoCedar + Kimi K3 & 100.00 & 100.00 & 100.00 & Same first query as Direct, 2 candidates\\[4pt]

AutoCedar + Kimi K2.6 & 100.00 & 100.00 & 100.00 & Same first query as Direct, 2 candidates\\[4pt]

\midrule

\multicolumn{5}{l}{\textit{Recorded intermediate representations}}\\[3pt]

\multicolumn{5}{@{}p{0.98\textwidth}@{}}{\textbf{Roundtrip.} Reconstructed reading Retrieves the artist\_name (aliased as name), country, and age for every row in the singer table, with no filtering conditions, no aggregation, and no exclusions. The result rows are ordered by age in descending order.}\\[4pt]

\multicolumn{5}{@{}p{0.98\textwidth}@{}}{\textbf{Clover.} SQL reconstruction and specification checks matched. The reconstructed description was assessed by the model.}\\[4pt]

\bottomrule

\end{tabular}

\end{table*}

\begin{table*}[p]

\centering\footnotesize

\setlength{\tabcolsep}{3pt}

\caption{Workflow comparison. Ambiguous input. Precision, recall, and F1 are percentages computed from the complete returned set against the benchmark references.}

\label{tab:sample-evidence-7}
\begin{tabular}{@{}p{0.19\textwidth}rrrp{0.57\textwidth}@{}}

\toprule

\multicolumn{5}{@{}p{0.98\textwidth}@{}}{\textbf{AMBIQT 000002}. Show name, country, age for all singers ordered by age from the oldest to the youngest.}\\[4pt]

\multicolumn{5}{@{}p{0.98\textwidth}@{}}{\textit{Reference SQL 1.} {\ttfamily\raggedright select name,\allowbreak{} nationality,\allowbreak{} age from singer order by age desc}\par\smallskip \textit{Reference SQL 2.} {\ttfamily\raggedright select name,\allowbreak{} citizenship,\allowbreak{} age from singer order by age desc}}\\[5pt]

Workflow & P & R & F1 & First returned SQL and candidate count\\

\midrule

Direct & 100.00 & 100.00 & 100.00 & {\ttfamily\raggedright select name,\allowbreak{} nationality,\allowbreak{} age from singer order by age desc}, 5 candidates\\[4pt]

AutoCedar & 100.00 & 100.00 & 100.00 & Same first query as Direct, 3 candidates\\[4pt]

Roundtrip & 100.00 & 100.00 & 100.00 & {\ttfamily\raggedright SELECT name,\allowbreak{} nationality AS country,\allowbreak{} age FROM singer ORDER BY age DESC}, 2 candidates\\[4pt]

Clover & 0.00 & 0.00 & 0.00 & Empty returned set, 0 candidates\\[4pt]

IntentPrompt & 100.00 & 100.00 & 100.00 & Same first query as Roundtrip, 2 candidates\\[4pt]

AutoCedar + Kimi K3 & 66.67 & 100.00 & 80.00 & Same first query as Direct, 3 candidates\\[4pt]

AutoCedar + Kimi K2.6 & 100.00 & 100.00 & 100.00 & Same first query as Direct, 2 candidates\\[4pt]

\midrule

\multicolumn{5}{l}{\textit{Recorded intermediate representations}}\\[3pt]

\multicolumn{5}{@{}p{0.98\textwidth}@{}}{\textbf{Roundtrip.} Reconstructed reading Retrieve the name, nationality (renamed as country), and age of every row in the singer table, with no filtering or aggregation, ordered by age from highest to lowest.}\\[4pt]

\multicolumn{5}{@{}p{0.98\textwidth}@{}}{\textbf{Clover.} SQL reconstruction and specification checks mismatched. The reconstructed description was assessed by the model.}\\[4pt]

\bottomrule

\end{tabular}

\end{table*}

\begin{table*}[p]

\centering\footnotesize

\setlength{\tabcolsep}{3pt}

\caption{Workflow comparison. Ambiguous input. Precision, recall, and F1 are percentages computed from the complete returned set against the benchmark references.}

\label{tab:sample-evidence-8}
\begin{tabular}{@{}p{0.19\textwidth}rrrp{0.57\textwidth}@{}}

\toprule

\multicolumn{5}{@{}p{0.98\textwidth}@{}}{\textbf{AMBIQT 000003}. What are the names, countries, and ages for every singer in descending order of age.}\\[4pt]

\multicolumn{5}{@{}p{0.98\textwidth}@{}}{\textit{Reference SQL 1.} {\ttfamily\raggedright select artist\_\allowbreak{}name,\allowbreak{} country,\allowbreak{} age from singer order by age desc}\par\smallskip \textit{Reference SQL 2.} {\ttfamily\raggedright select performer\_\allowbreak{}name,\allowbreak{} country,\allowbreak{} age from singer order by age desc}}\\[5pt]

Workflow & P & R & F1 & First returned SQL and candidate count\\

\midrule

Direct & 66.67 & 100.00 & 80.00 & {\ttfamily\raggedright select performer\_\allowbreak{}name,\allowbreak{} country,\allowbreak{} age from singer order by age desc}, 5 candidates\\[4pt]

AutoCedar & 66.67 & 100.00 & 80.00 & Same first query as Direct, 3 candidates\\[4pt]

Roundtrip & 100.00 & 100.00 & 100.00 & {\ttfamily\raggedright SELECT performer\_\allowbreak{}name AS name,\allowbreak{} country,\allowbreak{} age FROM singer ORDER BY age DESC}, 2 candidates\\[4pt]

Clover & 66.67 & 100.00 & 80.00 & Same first query as Direct, 3 candidates\\[4pt]

IntentPrompt & 66.67 & 100.00 & 80.00 & {\ttfamily\raggedright SELECT artist\_\allowbreak{}name,\allowbreak{} country,\allowbreak{} age FROM singer ORDER BY age DESC}, 3 candidates\\[4pt]

AutoCedar + Kimi K3 & 66.67 & 100.00 & 80.00 & Same first query as Direct, 3 candidates\\[4pt]

AutoCedar + Kimi K2.6 & 100.00 & 100.00 & 100.00 & Same first query as Direct, 2 candidates\\[4pt]

\midrule

\multicolumn{5}{l}{\textit{Recorded intermediate representations}}\\[3pt]

\multicolumn{5}{@{}p{0.98\textwidth}@{}}{\textbf{Roundtrip.} Reconstructed reading For every row in the singer table, return the performer\_name value under the alias name, along with country and age. The result includes all rows with no filtering or exclusions and no aggregation. The rows are (excerpt)}\\[4pt]

\multicolumn{5}{@{}p{0.98\textwidth}@{}}{\textbf{Clover.} SQL reconstruction and specification checks matched. The reconstructed description was assessed by the model.}\\[4pt]

\bottomrule

\end{tabular}

\end{table*}

\begin{table*}[p]

\centering\footnotesize

\setlength{\tabcolsep}{3pt}

\caption{Workflow comparison. Ambiguous input. Precision, recall, and F1 are percentages computed from the complete returned set against the benchmark references.}

\label{tab:sample-evidence-9}
\begin{tabular}{@{}p{0.19\textwidth}rrrp{0.57\textwidth}@{}}

\toprule

\multicolumn{5}{@{}p{0.98\textwidth}@{}}{\textbf{AMBIQT 000004}. What are the names, countries, and ages for every singer in descending order of age.}\\[4pt]

\multicolumn{5}{@{}p{0.98\textwidth}@{}}{\textit{Reference SQL 1.} {\ttfamily\raggedright select name,\allowbreak{} country,\allowbreak{} years\_\allowbreak{}old from singer order by years\_\allowbreak{}old desc}\par\smallskip \textit{Reference SQL 2.} {\ttfamily\raggedright select name,\allowbreak{} country,\allowbreak{} birthdate from singer order by birthdate desc}}\\[5pt]

Workflow & P & R & F1 & First returned SQL and candidate count\\

\midrule

Direct & 100.00 & 100.00 & 100.00 & {\ttfamily\raggedright select name,\allowbreak{} country,\allowbreak{} years\_\allowbreak{}old from singer order by years\_\allowbreak{}old desc}, 5 candidates\\[4pt]

AutoCedar & 100.00 & 50.00 & 66.67 & Same first query as Direct, 2 candidates\\[4pt]

Roundtrip & 100.00 & 50.00 & 66.67 & Same first query as Direct, 1 candidate\\[4pt]

Clover & 0.00 & 0.00 & 0.00 & {\ttfamily\raggedright SELECT name,\allowbreak{} country,\allowbreak{} years\_\allowbreak{}old FROM singer ORDER BY years\_\allowbreak{}old ASC}, 1 candidate\\[4pt]

IntentPrompt & 100.00 & 50.00 & 66.67 & {\ttfamily\raggedright SELECT name,\allowbreak{} country,\allowbreak{} years\_\allowbreak{}old AS age FROM singer ORDER BY years\_\allowbreak{}old DESC}, 1 candidate\\[4pt]

AutoCedar + Kimi K3 & 100.00 & 50.00 & 66.67 & Same first query as Direct, 2 candidates\\[4pt]

AutoCedar + Kimi K2.6 & 100.00 & 50.00 & 66.67 & Same first query as Direct, 2 candidates\\[4pt]

\midrule

\multicolumn{5}{l}{\textit{Recorded intermediate representations}}\\[3pt]

\multicolumn{5}{@{}p{0.98\textwidth}@{}}{\textbf{Roundtrip.} Reconstructed reading For every row in the singer table, return the singer's name, country, and years\_old. The result rows are ordered by years\_old from highest to lowest. There are no filters, exclusions, aggregations, or grouping.}\\[4pt]

\multicolumn{5}{@{}p{0.98\textwidth}@{}}{\textbf{Clover.} SQL reconstruction and specification checks matched. The reconstructed description was assessed by the model.}\\[4pt]

\bottomrule

\end{tabular}

\end{table*}

\begin{table*}[p]

\centering\footnotesize

\setlength{\tabcolsep}{3pt}

\caption{Workflow comparison. Ambiguous input. Precision, recall, and F1 are percentages computed from the complete returned set against the benchmark references.}

\label{tab:sample-evidence-10}
\begin{tabular}{@{}p{0.19\textwidth}rrrp{0.57\textwidth}@{}}

\toprule

\multicolumn{5}{@{}p{0.98\textwidth}@{}}{\textbf{AMBIQT 000005}. What are the names, countries, and ages for every singer in descending order of age.}\\[4pt]

\multicolumn{5}{@{}p{0.98\textwidth}@{}}{\textit{Reference SQL 1.} {\ttfamily\raggedright select name,\allowbreak{} nationality,\allowbreak{} age from singer order by age desc}\par\smallskip \textit{Reference SQL 2.} {\ttfamily\raggedright select name,\allowbreak{} citizenship,\allowbreak{} age from singer order by age desc}}\\[5pt]

Workflow & P & R & F1 & First returned SQL and candidate count\\

\midrule

Direct & 66.67 & 100.00 & 80.00 & {\ttfamily\raggedright select name,\allowbreak{} nationality,\allowbreak{} age from singer order by age desc}, 5 candidates\\[4pt]

AutoCedar & 66.67 & 100.00 & 80.00 & Same first query as Direct, 5 candidates\\[4pt]

Roundtrip & 100.00 & 100.00 & 100.00 & Same first query as Direct, 2 candidates\\[4pt]

Clover & 0.00 & 0.00 & 0.00 & {\ttfamily\raggedright SELECT name,\allowbreak{} nationality,\allowbreak{} citizenship,\allowbreak{} age FROM singer ORDER BY CAST(age AS INTEGER) DESC}, 1 candidate\\[4pt]

IntentPrompt & 66.67 & 100.00 & 80.00 & Same first query as Direct, 3 candidates\\[4pt]

AutoCedar + Kimi K3 & 66.67 & 100.00 & 80.00 & Same first query as Direct, 5 candidates\\[4pt]

AutoCedar + Kimi K2.6 & 66.67 & 100.00 & 80.00 & Same first query as Direct, 4 candidates\\[4pt]

\midrule

\multicolumn{5}{l}{\textit{Recorded intermediate representations}}\\[3pt]

\multicolumn{5}{@{}p{0.98\textwidth}@{}}{\textbf{Roundtrip.} Reconstructed reading Retrieve the name, nationality, and age for every row in the singer table, with the results ordered by age from highest to lowest. No rows are excluded and no aggregation is performed.}\\[4pt]

\multicolumn{5}{@{}p{0.98\textwidth}@{}}{\textbf{Clover.} SQL reconstruction and specification checks matched. The reconstructed description was assessed by the model.}\\[4pt]

\bottomrule

\end{tabular}

\end{table*}

\begin{table*}[p]

\centering\footnotesize

\setlength{\tabcolsep}{3pt}

\caption{Interpretation before AutoCedar. Clear input. Precision, recall, and F1 are percentages computed from the complete returned set against the benchmark references.}

\label{tab:sample-evidence-11}
\begin{tabular}{@{}p{0.19\textwidth}rrrp{0.57\textwidth}@{}}

\toprule

\multicolumn{5}{@{}p{0.98\textwidth}@{}}{\textbf{AMBROSIA 1481}. Please display the name and description of programs that are consistent across all authorities}\\[4pt]

\multicolumn{5}{@{}p{0.98\textwidth}@{}}{\textit{Reference SQL.} {\ttfamily\raggedright SELECT Programs.\allowbreak{}program\_\allowbreak{}name,\allowbreak{} Programs.\allowbreak{}description from Authorities\_\allowbreak{}Programs join Programs on Authorities\_\allowbreak{}Programs.\allowbreak{}program\_\allowbreak{}id = Programs.\allowbreak{}program\_\allowbreak{}id join Authorities on Authorities\_\allowbreak{}Programs.\allowbreak{}authority\_\allowbreak{}id = Authorities.\allowbreak{}authority\_\allowbreak{}id GROUP BY Authorities\_\allowbreak{}Programs.\allowbreak{}program\_\allowbreak{}id HAVING COUNT(DISTINCT Authorities\_\allowbreak{}Programs.\allowbreak{}authority\_\allowbreak{}id) = (SELECT COUNT(*) FROM Authorities)}}\\[5pt]

Workflow & P & R & F1 & First returned SQL and candidate count\\

\midrule

AutoCedar & 100.00 & 100.00 & 100.00 & {\ttfamily\raggedright SELECT p.\allowbreak{}program\_\allowbreak{}name,\allowbreak{} p.\allowbreak{}description FROM Programs p WHERE NOT EXISTS (SELECT 1 FROM Authorities a WHERE NOT EXISTS (SELECT 1 FROM Authorities\_\allowbreak{}Programs ap WHERE ap.\allowbreak{}authority\_\allowbreak{}id = a.\allowbreak{}authority\_\allowbreak{}id AND ap.\allowbreak{}program\_\allowbreak{}id = p.\allowbreak{}program\_\allowbreak{}id))}, 2 candidates\\[4pt]

Free interpretation & 100.00 & 100.00 & 100.00 & {\ttfamily\raggedright SELECT p.\allowbreak{}program\_\allowbreak{}name AS name,\allowbreak{} p.\allowbreak{}description FROM Programs p WHERE NOT EXISTS (SELECT 1 FROM Authorities a WHERE NOT EXISTS (SELECT 1 FROM Authorities\_\allowbreak{}Programs ap WHERE ap.\allowbreak{}authority\_\allowbreak{}id = a.\allowbreak{}authority\_\allowbreak{}id AND ap.\allowbreak{}program\_\allowbreak{}id = p.\allowbreak{}program\_\allowbreak{}id))}, 6 candidates\\[4pt]

Structured interpretation & 100.00 & 100.00 & 100.00 & {\ttfamily\raggedright SELECT p.\allowbreak{}program\_\allowbreak{}name,\allowbreak{} p.\allowbreak{}description FROM Programs p WHERE NOT EXISTS ( SELECT 1 FROM Authorities a WHERE NOT EXISTS ( SELECT 1 FROM Authorities\_\allowbreak{}Programs ap WHERE ap.\allowbreak{}authority\_\allowbreak{}id = a.\allowbreak{}authority\_\allowbreak{}id AND ap.\allowbreak{}program\_\allowbreak{}id = p.\allowbreak{}program\_\allowbreak{}id ) )}, 3 candidates\\[4pt]

\midrule

\multicolumn{5}{l}{\textit{Recorded intermediate representations}}\\[3pt]

\multicolumn{5}{@{}p{0.98\textwidth}@{}}{\textbf{AutoCedar.} Atoms a1 = EXISTS ( SELECT 1 FROM AUTHORITIES A WHERE NOT EXISTS ( SELECT 1 FROM AUTHORITIES\_PROGRAMS AP WHERE AP . AUTHORITY\_ID = A . AUTHORITY\_ID AND AP . PROGRAM\_ID = P . PROGRAM\_ID ) ). Boolean form NOT (a1)}\\[4pt]

\multicolumn{5}{@{}p{0.98\textwidth}@{}}{\textbf{Free interpretation.} Reading A program is consistent across all authorities if every authority in the Authorities table is linked to that program, output the program's name and description.}\\[4pt]

\multicolumn{5}{@{}p{0.98\textwidth}@{}}{\textbf{Structured interpretation.} Reading Display the program\_name and description of each program that is associated with every authority in the Authorities table, 'consistent across all authorities' is read as the program being common/available across all authorities.}\\[4pt]

\bottomrule

\end{tabular}

\end{table*}

\begin{table*}[p]

\centering\footnotesize

\setlength{\tabcolsep}{3pt}

\caption{Interpretation before AutoCedar. Clear input. Precision, recall, and F1 are percentages computed from the complete returned set against the benchmark references.}

\label{tab:sample-evidence-12}
\begin{tabular}{@{}p{0.19\textwidth}rrrp{0.57\textwidth}@{}}

\toprule

\multicolumn{5}{@{}p{0.98\textwidth}@{}}{\textbf{AMBROSIA 1826}. Which fashion designer brings all the reporters together.}\\[4pt]

\multicolumn{5}{@{}p{0.98\textwidth}@{}}{\textit{Reference SQL.} {\ttfamily\raggedright SELECT Designers.\allowbreak{}Name from Fashion\_\allowbreak{}Reporters\_\allowbreak{}Designers join Designers on Fashion\_\allowbreak{}Reporters\_\allowbreak{}Designers.\allowbreak{}DesignerID = Designers.\allowbreak{}DesignerID join Fashion\_\allowbreak{}Reporters on Fashion\_\allowbreak{}Reporters\_\allowbreak{}Designers.\allowbreak{}ReporterID = Fashion\_\allowbreak{}Reporters.\allowbreak{}ReporterID GROUP BY Fashion\_\allowbreak{}Reporters\_\allowbreak{}Designers.\allowbreak{}DesignerID HAVING COUNT(DISTINCT Fashion\_\allowbreak{}Reporters\_\allowbreak{}Designers.\allowbreak{}ReporterID) = (SELECT COUNT(*) FROM Fashion\_\allowbreak{}Reporters)}}\\[5pt]

Workflow & P & R & F1 & First returned SQL and candidate count\\

\midrule

AutoCedar & 100.00 & 100.00 & 100.00 & {\ttfamily\raggedright SELECT D.\allowbreak{}Name FROM Designers D JOIN Fashion\_\allowbreak{}Reporters\_\allowbreak{}Designers FRD ON D.\allowbreak{}DesignerID = FRD.\allowbreak{}DesignerID GROUP BY D.\allowbreak{}DesignerID,\allowbreak{} D.\allowbreak{}Name HAVING COUNT(DISTINCT FRD.\allowbreak{}ReporterID) = (SELECT COUNT(*) FROM Fashion\_\allowbreak{}Reporters)}, 2 candidates\\[4pt]

Free interpretation & 100.00 & 100.00 & 100.00 & {\ttfamily\raggedright SELECT d.\allowbreak{}Name FROM Designers d WHERE NOT EXISTS (SELECT 1 FROM Fashion\_\allowbreak{}Reporters r WHERE NOT EXISTS (SELECT 1 FROM Fashion\_\allowbreak{}Reporters\_\allowbreak{}Designers frd WHERE frd.\allowbreak{}DesignerID = d.\allowbreak{}DesignerID AND frd.\allowbreak{}ReporterID = r.\allowbreak{}ReporterID))}, 5 candidates\\[4pt]

Structured interpretation & 100.00 & 100.00 & 100.00 & {\ttfamily\raggedright SELECT d.\allowbreak{}Name FROM Designers d WHERE NOT EXISTS (SELECT 1 FROM Fashion\_\allowbreak{}Reporters r WHERE NOT EXISTS (SELECT 1 FROM Fashion\_\allowbreak{}Reporters\_\allowbreak{}Designers rd WHERE rd.\allowbreak{}DesignerID = d.\allowbreak{}DesignerID AND rd.\allowbreak{}ReporterID = r.\allowbreak{}ReporterID))}, 1 candidate\\[4pt]

\midrule

\multicolumn{5}{l}{\textit{Recorded intermediate representations}}\\[3pt]

\multicolumn{5}{@{}p{0.98\textwidth}@{}}{\textbf{AutoCedar.} Atoms a1 = D . DESIGNERID = FRD . DESIGNERID. Boolean form a1}\\[4pt]

\multicolumn{5}{@{}p{0.98\textwidth}@{}}{\textbf{Free interpretation.} Reading Designer associated with every reporter in Fashion\_Reporters.}\\[4pt]

\multicolumn{5}{@{}p{0.98\textwidth}@{}}{\textbf{Structured interpretation.} Reading Find the fashion designer who is associated with every reporter in the database, where the association is represented by Fashion\_Reporters\_Designers, return the designer's name.}\\[4pt]

\bottomrule

\end{tabular}

\end{table*}

\begin{table*}[p]

\centering\footnotesize

\setlength{\tabcolsep}{3pt}

\caption{Interpretation before AutoCedar. Clear input. Precision, recall, and F1 are percentages computed from the complete returned set against the benchmark references.}

\label{tab:sample-evidence-13}
\begin{tabular}{@{}p{0.19\textwidth}rrrp{0.57\textwidth}@{}}

\toprule

\multicolumn{5}{@{}p{0.98\textwidth}@{}}{\textbf{AMBROSIA 2666}. What is the version of program common for all visual effects artists.}\\[4pt]

\multicolumn{5}{@{}p{0.98\textwidth}@{}}{\textit{Reference SQL.} {\ttfamily\raggedright SELECT Programs.\allowbreak{}version from Visual\_\allowbreak{}Effects\_\allowbreak{}Artists\_\allowbreak{}Programs join Programs on Visual\_\allowbreak{}Effects\_\allowbreak{}Artists\_\allowbreak{}Programs.\allowbreak{}program\_\allowbreak{}id = Programs.\allowbreak{}program\_\allowbreak{}id join Visual\_\allowbreak{}Effects\_\allowbreak{}Artists on Visual\_\allowbreak{}Effects\_\allowbreak{}Artists\_\allowbreak{}Programs.\allowbreak{}visual\_\allowbreak{}effects\_\allowbreak{}artist\_\allowbreak{}id = Visual\_\allowbreak{}Effects\_\allowbreak{}Artists.\allowbreak{}artist\_\allowbreak{}id GROUP BY Visual\_\allowbreak{}Effects\_\allowbreak{}Artists\_\allowbreak{}Programs.\allowbreak{}program\_\allowbreak{}id HAVING COUNT(DISTINCT Visual\_\allowbreak{}Effects\_\allowbreak{}Artists\_\allowbreak{}Programs.\allowbreak{}visual\_\allowbreak{}effects\_\allowbreak{}artist\_\allowbreak{}id) = (SELECT COUNT(*) FROM Visual\_\allowbreak{}Effects\_\allowbreak{}Artists)}}\\[5pt]

Workflow & P & R & F1 & First returned SQL and candidate count\\

\midrule

AutoCedar & 100.00 & 100.00 & 100.00 & {\ttfamily\raggedright SELECT DISTINCT p.\allowbreak{}version FROM Programs p WHERE NOT EXISTS (SELECT 1 FROM Visual\_\allowbreak{}Effects\_\allowbreak{}Artists a WHERE NOT EXISTS (SELECT 1 FROM Visual\_\allowbreak{}Effects\_\allowbreak{}Artists\_\allowbreak{}Programs veap WHERE veap.\allowbreak{}visual\_\allowbreak{}effects\_\allowbreak{}artist\_\allowbreak{}id = a.\allowbreak{}artist\_\allowbreak{}id AND veap.\allowbreak{}program\_\allowbreak{}id = p.\allowbreak{}program\_\allowbreak{}id))}, 2 candidates\\[4pt]

Free interpretation & 100.00 & 100.00 & 100.00 & {\ttfamily\raggedright SELECT DISTINCT p.\allowbreak{}version FROM Programs AS p JOIN Visual\_\allowbreak{}Effects\_\allowbreak{}Artists\_\allowbreak{}Programs AS vap ON p.\allowbreak{}program\_\allowbreak{}id = vap.\allowbreak{}program\_\allowbreak{}id GROUP BY p.\allowbreak{}program\_\allowbreak{}id HAVING COUNT(DISTINCT vap.\allowbreak{}visual\_\allowbreak{}effects\_\allowbreak{}artist\_\allowbreak{}id) = (SELECT COUNT(*) FROM Visual\_\allowbreak{}Effects\_\allowbreak{}Artists)}, 12 candidates\\[4pt]

Structured interpretation & 100.00 & 100.00 & 100.00 & {\ttfamily\raggedright SELECT DISTINCT p.\allowbreak{}version FROM Programs p JOIN Visual\_\allowbreak{}Effects\_\allowbreak{}Artists\_\allowbreak{}Programs vap ON p.\allowbreak{}program\_\allowbreak{}id = vap.\allowbreak{}program\_\allowbreak{}id GROUP BY p.\allowbreak{}program\_\allowbreak{}id,\allowbreak{} p.\allowbreak{}version HAVING COUNT(DISTINCT vap.\allowbreak{}visual\_\allowbreak{}effects\_\allowbreak{}artist\_\allowbreak{}id) = (SELECT COUNT(*) FROM Visual\_\allowbreak{}Effects\_\allowbreak{}Artists)}, 7 candidates\\[4pt]

\midrule

\multicolumn{5}{l}{\textit{Recorded intermediate representations}}\\[3pt]

\multicolumn{5}{@{}p{0.98\textwidth}@{}}{\textbf{AutoCedar.} Atoms a1 = EXISTS ( SELECT 1 FROM VISUAL\_EFFECTS\_ARTISTS A WHERE NOT EXISTS ( SELECT 1 FROM VISUAL\_EFFECTS\_ARTISTS\_PROGRAMS VEAP WHERE VEAP . VISUAL\_EFFECTS\_ARTIST\_ID = A . ARTIST\_ID AND VEAP . PROGRAM\_ID = P . PROGRAM\_ID ) ). Boolean form (excerpt)}\\[4pt]

\multicolumn{5}{@{}p{0.98\textwidth}@{}}{\textbf{Free interpretation.} Reading Common means shared by every artist, program is identified by program\_id, return the version of each program that all artists use.}\\[4pt]

\multicolumn{5}{@{}p{0.98\textwidth}@{}}{\textbf{Structured interpretation.} Reading Find the version of a program that is associated with every visual effects artist, 'common' is read as universally shared/intersection, and 'the version' is the version of that shared program.}\\[4pt]

\bottomrule

\end{tabular}

\end{table*}

\begin{table*}[p]

\centering\footnotesize

\setlength{\tabcolsep}{3pt}

\caption{Interpretation before AutoCedar. Ambiguous input. Precision, recall, and F1 are percentages computed from the complete returned set against the benchmark references.}

\label{tab:sample-evidence-14}
\begin{tabular}{@{}p{0.19\textwidth}rrrp{0.57\textwidth}@{}}

\toprule

\multicolumn{5}{@{}p{0.98\textwidth}@{}}{\textbf{AMBROSIA 2941}. Who is the chef at the restaurant where you can find a grilled steak with garlic butter on the menu.}\\[4pt]

\multicolumn{5}{@{}p{0.98\textwidth}@{}}{\textit{Reference SQL 1.} {\ttfamily\raggedright SELECT r.\allowbreak{}executive\_\allowbreak{}chef FROM Restaurants r,\allowbreak{} MenuItems m WHERE r.\allowbreak{}restaurant\_\allowbreak{}id = m.\allowbreak{}restaurant\_\allowbreak{}id AND m.\allowbreak{}item\_\allowbreak{}name = 'Steak'}\par\smallskip \textit{Reference SQL 2.} {\ttfamily\raggedright SELECT r.\allowbreak{}sous\_\allowbreak{}chef FROM Restaurants r,\allowbreak{} MenuItems m WHERE r.\allowbreak{}restaurant\_\allowbreak{}id = m.\allowbreak{}restaurant\_\allowbreak{}id AND m.\allowbreak{}item\_\allowbreak{}name = 'Steak'}\par\smallskip \textit{Reference SQL 3.} {\ttfamily\raggedright SELECT r.\allowbreak{}executive\_\allowbreak{}chef,\allowbreak{} sous\_\allowbreak{}chef FROM Restaurants r,\allowbreak{} MenuItems m WHERE r.\allowbreak{}restaurant\_\allowbreak{}id = m.\allowbreak{}restaurant\_\allowbreak{}id AND m.\allowbreak{}item\_\allowbreak{}name = 'Steak'}}\\[5pt]

Workflow & P & R & F1 & First returned SQL and candidate count\\

\midrule

AutoCedar & 66.67 & 66.67 & 66.67 & {\ttfamily\raggedright SELECT R.\allowbreak{}executive\_\allowbreak{}chef FROM Restaurants R JOIN MenuItems M ON R.\allowbreak{}restaurant\_\allowbreak{}id = M.\allowbreak{}restaurant\_\allowbreak{}id WHERE M.\allowbreak{}description = 'Grilled steak with garlic butter'}, 3 candidates\\[4pt]

Free interpretation & 66.67 & 66.67 & 66.67 & {\ttfamily\raggedright SELECT DISTINCT r.\allowbreak{}executive\_\allowbreak{}chef AS chef FROM Restaurants r JOIN MenuItems m ON r.\allowbreak{}restaurant\_\allowbreak{}id = m.\allowbreak{}restaurant\_\allowbreak{}id WHERE LOWER(m.\allowbreak{}description) = 'grilled steak with garlic butter'}, 17 candidates\\[4pt]

Structured interpretation & 66.67 & 66.67 & 66.67 & {\ttfamily\raggedright SELECT r.\allowbreak{}executive\_\allowbreak{}chef AS chef FROM Restaurants r JOIN MenuItems m ON r.\allowbreak{}restaurant\_\allowbreak{}id = m.\allowbreak{}restaurant\_\allowbreak{}id WHERE m.\allowbreak{}description LIKE '\%grilled steak with garlic butter\%'}, 7 candidates\\[4pt]

\midrule

\multicolumn{5}{l}{\textit{Recorded intermediate representations}}\\[3pt]

\multicolumn{5}{@{}p{0.98\textwidth}@{}}{\textbf{AutoCedar.} Atoms a1 = R . RESTAURANT\_ID = M . RESTAURANT\_ID. Boolean form a1}\\[4pt]

\multicolumn{5}{@{}p{0.98\textwidth}@{}}{\textbf{Free interpretation.} Reading “The chef” refers to the executive chef, and the menu item is the one whose description exactly matches the phrase, ignoring the article “a” and treating case insensitively. Return distinct executive chefs for any matching (excerpt)}\\[4pt]

\multicolumn{5}{@{}p{0.98\textwidth}@{}}{\textbf{Structured interpretation.} Reading The question asks for the executive chef of the restaurant that offers a menu item matching 'grilled steak with garlic butter'. 'Chef' is read as the Restaurants.executive\_chef column.}\\[4pt]

\bottomrule

\end{tabular}

\end{table*}

\begin{table*}[p]

\centering\footnotesize

\setlength{\tabcolsep}{3pt}

\caption{Interpretation before AutoCedar. Clear input. Precision, recall, and F1 are percentages computed from the complete returned set against the benchmark references.}

\label{tab:sample-evidence-15}
\begin{tabular}{@{}p{0.19\textwidth}rrrp{0.57\textwidth}@{}}

\toprule

\multicolumn{5}{@{}p{0.98\textwidth}@{}}{\textbf{AMBROSIA 1130}. Provide a list of all baggage claim staff and also provide a list of security staff in terminal 2. Show them in one table.}\\[4pt]

\multicolumn{5}{@{}p{0.98\textwidth}@{}}{\textit{Reference SQL.} {\ttfamily\raggedright SELECT BaggageClaims.\allowbreak{}HandlerName as name FROM BaggageClaims UNION SELECT SecurityChecks.\allowbreak{}OfficerName as name FROM SecurityChecks WHERE SecurityChecks.\allowbreak{}Location = "Terminal 2"}}\\[5pt]

Workflow & P & R & F1 & First returned SQL and candidate count\\

\midrule

AutoCedar & 33.33 & 100.00 & 50.00 & {\ttfamily\raggedright SELECT HandlerName AS StaffName FROM BaggageClaims UNION SELECT OfficerName AS StaffName FROM SecurityChecks WHERE Location = 'Terminal 2'}, 4 candidates\\[4pt]

Free interpretation & 33.33 & 100.00 & 50.00 & {\ttfamily\raggedright SELECT HandlerName AS StaffName FROM BaggageClaims UNION ALL SELECT OfficerName FROM SecurityChecks WHERE Location = 'Terminal 2'}, 12 candidates\\[4pt]

Structured interpretation & 25.00 & 100.00 & 40.00 & {\ttfamily\raggedright SELECT HandlerName AS StaffName FROM BaggageClaims UNION ALL SELECT OfficerName AS StaffName FROM SecurityChecks WHERE Location = 'Terminal 2'}, 5 candidates\\[4pt]

\midrule

\multicolumn{5}{l}{\textit{Recorded intermediate representations}}\\[3pt]

\multicolumn{5}{@{}p{0.98\textwidth}@{}}{\textbf{AutoCedar.} Atoms a1 = LOCATION = 'Terminal 2'. Boolean form a1}\\[4pt]

\multicolumn{5}{@{}p{0.98\textwidth}@{}}{\textbf{Free interpretation.} Reading Stacked one-column list of names, all baggage claim handler rows and all security officer rows in Terminal 2, duplicates retained, no role column.}\\[4pt]

\multicolumn{5}{@{}p{0.98\textwidth}@{}}{\textbf{Structured interpretation.} Reading Root reading, the phrase 'in terminal 2' restricts only the security staff list, the baggage claim staff list is all baggage claim handlers regardless of terminal. The requested output is one table combining the two lists, with the (excerpt)}\\[4pt]

\bottomrule

\end{tabular}

\end{table*}

\begin{table*}[p]

\centering\footnotesize

\setlength{\tabcolsep}{3pt}

\caption{Interpretation before AutoCedar. Clear input. Precision, recall, and F1 are percentages computed from the complete returned set against the benchmark references.}

\label{tab:sample-evidence-16}
\begin{tabular}{@{}p{0.19\textwidth}rrrp{0.57\textwidth}@{}}

\toprule

\multicolumn{5}{@{}p{0.98\textwidth}@{}}{\textbf{AMBROSIA 1107}. Can you show both Boeing and Airbus aircraft models that have 180 seats. Show them in one table.}\\[4pt]

\multicolumn{5}{@{}p{0.98\textwidth}@{}}{\textit{Reference SQL.} {\ttfamily\raggedright SELECT Boeing.\allowbreak{}Model FROM Boeing WHERE Boeing.\allowbreak{}Number\_\allowbreak{}of\_\allowbreak{}seats = 180 UNION SELECT Airbus.\allowbreak{}Model FROM Airbus WHERE Airbus.\allowbreak{}Number\_\allowbreak{}of\_\allowbreak{}seats = 180}}\\[5pt]

Workflow & P & R & F1 & First returned SQL and candidate count\\

\midrule

AutoCedar & 100.00 & 100.00 & 100.00 & {\ttfamily\raggedright SELECT Model FROM Airbus WHERE Number\_\allowbreak{}of\_\allowbreak{}seats = 180 UNION SELECT Model FROM Boeing WHERE Number\_\allowbreak{}of\_\allowbreak{}seats = 180}, 2 candidates\\[4pt]

Free interpretation & 25.00 & 100.00 & 40.00 & Same first query as AutoCedar, 9 candidates\\[4pt]

Structured interpretation & 100.00 & 100.00 & 100.00 & Same first query as AutoCedar, 4 candidates\\[4pt]

\midrule

\multicolumn{5}{l}{\textit{Recorded intermediate representations}}\\[3pt]

\multicolumn{5}{@{}p{0.98\textwidth}@{}}{\textbf{AutoCedar.} Atoms a1 = NUMBER\_OF\_SEATS = 180. Boolean form a1}\\[4pt]

\multicolumn{5}{@{}p{0.98\textwidth}@{}}{\textbf{Free interpretation.} Reading Combine Airbus and Boeing models with exactly 180 seats using UNION, returning only distinct model names.}\\[4pt]

\multicolumn{5}{@{}p{0.98\textwidth}@{}}{\textbf{Structured interpretation.} Reading Interpret the request as asking for one combined result table of aircraft model names from the Airbus table and the Boeing table, restricted to entries whose seat count is 180, the manufacturer/source table is not part of the (excerpt)}\\[4pt]

\bottomrule

\end{tabular}

\end{table*}

\begin{table*}[p]

\centering\footnotesize

\setlength{\tabcolsep}{3pt}

\caption{Interpretation before AutoCedar. Clear input. Precision, recall, and F1 are percentages computed from the complete returned set against the benchmark references.}

\label{tab:sample-evidence-17}
\begin{tabular}{@{}p{0.19\textwidth}rrrp{0.57\textwidth}@{}}

\toprule

\multicolumn{5}{@{}p{0.98\textwidth}@{}}{\textbf{AMBROSIA 2584}. For all craft services, list their menus}\\[4pt]

\multicolumn{5}{@{}p{0.98\textwidth}@{}}{\textit{Reference SQL.} {\ttfamily\raggedright select craft\_\allowbreak{}services.\allowbreak{}description ,\allowbreak{} menus.\allowbreak{}name from Craft\_\allowbreak{}Services\_\allowbreak{}Menus join Menus on Craft\_\allowbreak{}Services\_\allowbreak{}Menus.\allowbreak{}MenuID = Menus.\allowbreak{}MenuID join Craft\_\allowbreak{}Services on Craft\_\allowbreak{}Services\_\allowbreak{}Menus.\allowbreak{}ServiceID = Craft\_\allowbreak{}Services.\allowbreak{}ServiceID}}\\[5pt]

Workflow & P & R & F1 & First returned SQL and candidate count\\

\midrule

AutoCedar & 8.33 & 100.00 & 15.38 & {\ttfamily\raggedright SELECT Craft\_\allowbreak{}Services.\allowbreak{}Description,\allowbreak{} Menus.\allowbreak{}Name FROM Craft\_\allowbreak{}Services JOIN Craft\_\allowbreak{}Services\_\allowbreak{}Menus ON Craft\_\allowbreak{}Services.\allowbreak{}ServiceID = Craft\_\allowbreak{}Services\_\allowbreak{}Menus.\allowbreak{}ServiceID JOIN Menus ON Craft\_\allowbreak{}Services\_\allowbreak{}Menus.\allowbreak{}MenuID = Menus.\allowbreak{}MenuID}, 23 candidates\\[4pt]

Free interpretation & 7.14 & 100.00 & 13.33 & {\ttfamily\raggedright SELECT cs.\allowbreak{}Description,\allowbreak{} m.\allowbreak{}Name FROM Craft\_\allowbreak{}Services cs JOIN Craft\_\allowbreak{}Services\_\allowbreak{}Menus csm ON cs.\allowbreak{}ServiceID = csm.\allowbreak{}ServiceID JOIN Menus m ON csm.\allowbreak{}MenuID = m.\allowbreak{}MenuID}, 21 candidates\\[4pt]

Structured interpretation & 12.50 & 100.00 & 22.22 & {\ttfamily\raggedright SELECT m.\allowbreak{}Name FROM Craft\_\allowbreak{}Services AS cs JOIN Craft\_\allowbreak{}Services\_\allowbreak{}Menus AS csm ON cs.\allowbreak{}ServiceID = csm.\allowbreak{}ServiceID JOIN Menus AS m ON csm.\allowbreak{}MenuID = m.\allowbreak{}MenuID}, 13 candidates\\[4pt]

\midrule

\multicolumn{5}{l}{\textit{Recorded intermediate representations}}\\[3pt]

\multicolumn{5}{@{}p{0.98\textwidth}@{}}{\textbf{AutoCedar.} Atoms a1 = CRAFT\_SERVICES . SERVICEID = CRAFT\_SERVICES\_MENUS . SERVICEID. Boolean form a1}\\[4pt]

\multicolumn{5}{@{}p{0.98\textwidth}@{}}{\textbf{Free interpretation.} Reading For each craft service, list each linked menu by service Description and menu Name.}\\[4pt]

\multicolumn{5}{@{}p{0.98\textwidth}@{}}{\textbf{Structured interpretation.} Reading General reading, For all craft services, retrieve the menu or menus associated with each service through the craft-service/menu mapping, the exact output grain, inclusion of services without menus, and column set are unspecified.}\\[4pt]

\bottomrule

\end{tabular}

\end{table*}

\begin{table*}[p]

\centering\footnotesize

\setlength{\tabcolsep}{3pt}

\caption{Interpretation before AutoCedar. Clear input. Precision, recall, and F1 are percentages computed from the complete returned set against the benchmark references.}

\label{tab:sample-evidence-18}
\begin{tabular}{@{}p{0.19\textwidth}rrrp{0.57\textwidth}@{}}

\toprule

\multicolumn{5}{@{}p{0.98\textwidth}@{}}{\textbf{AMBROSIA 2674}. What techniques are common to all sound designers.}\\[4pt]

\multicolumn{5}{@{}p{0.98\textwidth}@{}}{\textit{Reference SQL.} {\ttfamily\raggedright SELECT Techniques.\allowbreak{}MethodName from Sound\_\allowbreak{}Designer\_\allowbreak{}Technique join Techniques on Sound\_\allowbreak{}Designer\_\allowbreak{}Technique.\allowbreak{}TechniqueID = Techniques.\allowbreak{}ID join Sound\_\allowbreak{}Designers on Sound\_\allowbreak{}Designer\_\allowbreak{}Technique.\allowbreak{}SoundDesignerID = Sound\_\allowbreak{}Designers.\allowbreak{}ID GROUP BY Sound\_\allowbreak{}Designer\_\allowbreak{}Technique.\allowbreak{}TechniqueID HAVING COUNT(DISTINCT Sound\_\allowbreak{}Designer\_\allowbreak{}Technique.\allowbreak{}SoundDesignerID) = (SELECT COUNT(*) FROM Sound\_\allowbreak{}Designers)}}\\[5pt]

Workflow & P & R & F1 & First returned SQL and candidate count\\

\midrule

AutoCedar & 100.00 & 100.00 & 100.00 & {\ttfamily\raggedright SELECT T.\allowbreak{}MethodName FROM Techniques T WHERE NOT EXISTS (SELECT 1 FROM Sound\_\allowbreak{}Designers S WHERE NOT EXISTS (SELECT 1 FROM Sound\_\allowbreak{}Designer\_\allowbreak{}Technique SDT WHERE SDT.\allowbreak{}SoundDesignerID = S.\allowbreak{}ID AND SDT.\allowbreak{}TechniqueID = T.\allowbreak{}ID))}, 2 candidates\\[4pt]

Free interpretation & 50.00 & 100.00 & 66.67 & {\ttfamily\raggedright SELECT DISTINCT t.\allowbreak{}MethodName FROM Techniques t WHERE NOT EXISTS (SELECT 1 FROM Sound\_\allowbreak{}Designers sd WHERE NOT EXISTS (SELECT 1 FROM Sound\_\allowbreak{}Designer\_\allowbreak{}Technique sdt WHERE sdt.\allowbreak{}SoundDesignerID = sd.\allowbreak{}ID AND sdt.\allowbreak{}TechniqueID = t.\allowbreak{}ID))}, 6 candidates\\[4pt]

Structured interpretation & 100.00 & 100.00 & 100.00 & {\ttfamily\raggedright SELECT DISTINCT T.\allowbreak{}MethodName FROM Techniques T WHERE NOT EXISTS (SELECT 1 FROM Sound\_\allowbreak{}Designers S WHERE NOT EXISTS (SELECT 1 FROM Sound\_\allowbreak{}Designer\_\allowbreak{}Technique SDT WHERE SDT.\allowbreak{}SoundDesignerID = S.\allowbreak{}ID AND SDT.\allowbreak{}TechniqueID = T.\allowbreak{}ID))}, 1 candidate\\[4pt]

\midrule

\multicolumn{5}{l}{\textit{Recorded intermediate representations}}\\[3pt]

\multicolumn{5}{@{}p{0.98\textwidth}@{}}{\textbf{AutoCedar.} Atoms a1 = EXISTS ( SELECT 1 FROM SOUND\_DESIGNERS S WHERE NOT EXISTS ( SELECT 1 FROM SOUND\_DESIGNER\_TECHNIQUE SDT WHERE SDT . SOUNDDESIGNERID = S . ID AND SDT . TECHNIQUEID = T . ID ) ). Boolean form NOT (a1)}\\[4pt]

\multicolumn{5}{@{}p{0.98\textwidth}@{}}{\textbf{Free interpretation.} Reading Strict universal, a technique is common to all sound designers if it is linked to every row in Sound\_Designers, output method names.}\\[4pt]

\multicolumn{5}{@{}p{0.98\textwidth}@{}}{\textbf{Structured interpretation.} Reading Return the MethodName of each technique that is associated through Sound\_Designer\_Technique with every row in Sound\_Designers, a technique qualifies only if no sound designer in the database lacks an association with it.}\\[4pt]

\bottomrule

\end{tabular}

\end{table*}

\begin{table*}[p]

\centering\footnotesize

\setlength{\tabcolsep}{3pt}

\caption{Interpretation before AutoCedar. Clear input. Precision, recall, and F1 are percentages computed from the complete returned set against the benchmark references.}

\label{tab:sample-evidence-19}
\begin{tabular}{@{}p{0.19\textwidth}rrrp{0.57\textwidth}@{}}

\toprule

\multicolumn{5}{@{}p{0.98\textwidth}@{}}{\textbf{AMBROSIA 2976}. List drugs, their side effects and effectiveness in order of their popularity in prescriptions}\\[4pt]

\multicolumn{5}{@{}p{0.98\textwidth}@{}}{\textit{Reference SQL.} {\ttfamily\raggedright Select Drug.\allowbreak{}Name,\allowbreak{} Side\_\allowbreak{}Effects.\allowbreak{}EffectText,\allowbreak{} Side\_\allowbreak{}Effects.\allowbreak{}Severity,\allowbreak{} Effectiveness.\allowbreak{}percentage FROM Side\_\allowbreak{}Effects INNER JOIN Drug On Side\_\allowbreak{}Effects.\allowbreak{}sideEffectId = Drug.\allowbreak{}SideEffectId JOIN Prescription ON Prescription.\allowbreak{}drugID = Drug.\allowbreak{}drugID INNER JOIN Effectiveness On Effectiveness.\allowbreak{}EffectivenessId = Drug.\allowbreak{}EffectivenessId ORDER BY quantity}}\\[5pt]

Workflow & P & R & F1 & First returned SQL and candidate count\\

\midrule

AutoCedar & 0.00 & 0.00 & 0.00 & {\ttfamily\raggedright SELECT d.\allowbreak{}name,\allowbreak{} s.\allowbreak{}effectText,\allowbreak{} e.\allowbreak{}percentage FROM Drug d JOIN Side\_\allowbreak{}Effects s ON d.\allowbreak{}sideEffectID = s.\allowbreak{}sideEffectID JOIN Effectiveness e ON d.\allowbreak{}effectivenessID = e.\allowbreak{}effectivenessID LEFT JOIN Prescription p ON d.\allowbreak{}drugID = p.\allowbreak{}drugID GROUP BY d.\allowbreak{}drugID,\allowbreak{} d.\allowbreak{}name,\allowbreak{} s.\allowbreak{}effectText,\allowbreak{} e.\allowbreak{}percentage ORDER BY COUNT(p.\allowbreak{}prescriptionID) DESC}, 6 candidates\\[4pt]

Free interpretation & 0.00 & 0.00 & 0.00 & {\ttfamily\raggedright SELECT d.\allowbreak{}name,\allowbreak{} s.\allowbreak{}effectText,\allowbreak{} e.\allowbreak{}percentage FROM Drug d JOIN Prescription p ON p.\allowbreak{}drugID = d.\allowbreak{}drugID JOIN Side\_\allowbreak{}Effects s ON s.\allowbreak{}sideEffectID = d.\allowbreak{}sideEffectID JOIN Effectiveness e ON e.\allowbreak{}effectivenessID = d.\allowbreak{}effectivenessID GROUP BY d.\allowbreak{}drugID,\allowbreak{} d.\allowbreak{}name,\allowbreak{} s.\allowbreak{}effectText,\allowbreak{} e.\allowbreak{}percentage ORDER BY COUNT(p.\allowbreak{}prescriptionID) DESC}, 14 candidates\\[4pt]

Structured interpretation & 0.00 & 0.00 & 0.00 & {\ttfamily\raggedright SELECT d.\allowbreak{}name,\allowbreak{} se.\allowbreak{}effectText,\allowbreak{} e.\allowbreak{}percentage FROM Drug d JOIN Side\_\allowbreak{}Effects se ON d.\allowbreak{}sideEffectID = se.\allowbreak{}sideEffectID JOIN Effectiveness e ON d.\allowbreak{}effectivenessID = e.\allowbreak{}effectivenessID JOIN Prescription p ON d.\allowbreak{}drugID = p.\allowbreak{}drugID GROUP BY d.\allowbreak{}drugID,\allowbreak{} d.\allowbreak{}name,\allowbreak{} se.\allowbreak{}effectText,\allowbreak{} e.\allowbreak{}percentage ORDER BY COUNT(p.\allowbreak{}prescriptionID) DESC}, 19 candidates\\[4pt]

\midrule

\multicolumn{5}{l}{\textit{Recorded intermediate representations}}\\[3pt]

\multicolumn{5}{@{}p{0.98\textwidth}@{}}{\textbf{AutoCedar.} Atoms a1 = D . SIDEEFFECTID = S . SIDEEFFECTID. Boolean form a1}\\[4pt]

\multicolumn{5}{@{}p{0.98\textwidth}@{}}{\textbf{Free interpretation.} Reading List each prescribed drug by name with its side-effect text and effectiveness percentage, ordered by the number of prescription records for that drug, most prescriptions first, drugs without a matching side-effect row are omitted.}\\[4pt]

\multicolumn{5}{@{}p{0.98\textwidth}@{}}{\textbf{Structured interpretation.} Reading Popularity in prescriptions is the number of prescription records for each drug. List each drug with its side effect and effectiveness, ordered by that prescription count from most to least. Nested alternatives cover whether drugs (excerpt)}\\[4pt]

\bottomrule

\end{tabular}

\end{table*}

\begin{table*}[p]

\centering\footnotesize

\setlength{\tabcolsep}{3pt}

\caption{Interpretation before AutoCedar. Clear input. Precision, recall, and F1 are percentages computed from the complete returned set against the benchmark references.}

\label{tab:sample-evidence-20}
\begin{tabular}{@{}p{0.19\textwidth}rrrp{0.57\textwidth}@{}}

\toprule

\multicolumn{5}{@{}p{0.98\textwidth}@{}}{\textbf{AMBROSIA 3435}. Show both the government agencies and regulatory boards and the related engineering projects.}\\[4pt]

\multicolumn{5}{@{}p{0.98\textwidth}@{}}{\textit{Reference SQL.} {\ttfamily\raggedright select GovernmentAgency,\allowbreak{} RegulatoryBoard,\allowbreak{} EngineeringProject from Boards}}\\[5pt]

Workflow & P & R & F1 & First returned SQL and candidate count\\

\midrule

AutoCedar & 100.00 & 100.00 & 100.00 & {\ttfamily\raggedright SELECT GovernmentAgency,\allowbreak{} RegulatoryBoard,\allowbreak{} EngineeringProject FROM Boards}, 1 candidate\\[4pt]

Free interpretation & 100.00 & 100.00 & 100.00 & Same first query as AutoCedar, 2 candidates\\[4pt]

Structured interpretation & 50.00 & 100.00 & 66.67 & Same first query as AutoCedar, 2 candidates\\[4pt]

\midrule

\multicolumn{5}{l}{\textit{Recorded intermediate representations}}\\[3pt]

\multicolumn{5}{@{}p{0.98\textwidth}@{}}{\textbf{AutoCedar.} Reading Show the GovernmentAgency and RegulatoryBoard columns along with the related EngineeringProject column from the Boards table.}\\[4pt]

\multicolumn{5}{@{}p{0.98\textwidth}@{}}{\textbf{Free interpretation.} Reading Show the government agency, regulatory board, and engineering project from each row of the Boards table, preserving all rows and duplicates.}\\[4pt]

\multicolumn{5}{@{}p{0.98\textwidth}@{}}{\textbf{Structured interpretation.} Reading Return the government agency, regulatory board, and engineering project recorded together on each Boards row, so that each displayed project is the one related to that row's agency and board.}\\[4pt]

\bottomrule

\end{tabular}

\end{table*}

\begin{table*}[p]

\centering\footnotesize

\setlength{\tabcolsep}{3pt}

\caption{Confirmation feedback. Ambiguous input. Precision, recall, and F1 are percentages computed from the complete returned set against the benchmark references.}

\label{tab:sample-evidence-21}
\begin{tabular}{@{}p{0.19\textwidth}rrrp{0.57\textwidth}@{}}

\toprule

\multicolumn{5}{@{}p{0.98\textwidth}@{}}{\textbf{AMBROSIA 741}. What jobs are available for economists and statisticians working with demographic data. Show them in one table.}\\[4pt]

\multicolumn{5}{@{}p{0.98\textwidth}@{}}{\textit{Reference SQL 1.} {\ttfamily\raggedright SELECT JobPostings.\allowbreak{}title FROM JobPostings WHERE (JobPostings.\allowbreak{}Roles = "Economist" OR JobPostings.\allowbreak{}Roles = "Statistician") AND JobPostings.\allowbreak{}Data = "Demographics"}\par\smallskip \textit{Reference SQL 2.} {\ttfamily\raggedright SELECT JobPostings.\allowbreak{}title FROM JobPostings WHERE JobPostings.\allowbreak{}Roles = "Economist" OR JobPostings.\allowbreak{}Roles = "Statistician" AND JobPostings.\allowbreak{}Data = "Demographics"}}\\[5pt]

Workflow & P & R & F1 & First returned SQL and candidate count\\

\midrule

No extra model & 100.00 & 50.00 & 66.67 & {\ttfamily\raggedright SELECT Title FROM JobPostings WHERE (Roles = 'Economist' OR Roles = 'Statistician') AND Data = 'Demographics'}, 4 candidates\\[4pt]

Kimi K3 & 100.00 & 50.00 & 66.67 & Same first query as No extra model, 4 candidates\\[4pt]

Kimi K2.6 & 100.00 & 50.00 & 66.67 & Same first query as No extra model, 5 candidates\\[4pt]

\midrule

\multicolumn{5}{l}{\textit{Recorded intermediate representations}}\\[3pt]

\multicolumn{5}{@{}p{0.98\textwidth}@{}}{\textbf{No extra model.} Atoms a1 = DATA = 'Demographics', a2 = ROLES = 'Economist', a3 = ROLES = 'Statistician'. Boolean form ((a2 OR a3) AND a1)}\\[4pt]

\multicolumn{5}{@{}p{0.98\textwidth}@{}}{\textbf{Kimi K3.} Atoms a1 = DATA = 'Demographics', a2 = ROLES = 'Economist', a3 = ROLES = 'Statistician'. Boolean form ((a2 OR a3) AND a1)}\\[4pt]

\multicolumn{5}{@{}p{0.98\textwidth}@{}}{\textbf{Kimi K2.6.} Atoms a1 = DATA = 'Demographics', a2 = ROLES = 'Economist', a3 = ROLES = 'Statistician'. Boolean form ((a2 OR a3) AND a1)}\\[4pt]

\bottomrule

\end{tabular}

\end{table*}

\begin{table*}[p]

\centering\footnotesize

\setlength{\tabcolsep}{3pt}

\caption{Confirmation feedback. Clear input. Precision, recall, and F1 are percentages computed from the complete returned set against the benchmark references.}

\label{tab:sample-evidence-22}
\begin{tabular}{@{}p{0.19\textwidth}rrrp{0.57\textwidth}@{}}

\toprule

\multicolumn{5}{@{}p{0.98\textwidth}@{}}{\textbf{AMBROSIA 2002}. For every district, list the description of school grades.}\\[4pt]

\multicolumn{5}{@{}p{0.98\textwidth}@{}}{\textit{Reference SQL.} {\ttfamily\raggedright select district.\allowbreak{}name ,\allowbreak{} school\_\allowbreak{}grades.\allowbreak{}description from District\_\allowbreak{}School\_\allowbreak{}Grades join School\_\allowbreak{}Grades on District\_\allowbreak{}School\_\allowbreak{}Grades.\allowbreak{}school\_\allowbreak{}grade\_\allowbreak{}id = School\_\allowbreak{}Grades.\allowbreak{}id join District on District\_\allowbreak{}School\_\allowbreak{}Grades.\allowbreak{}district\_\allowbreak{}id = District.\allowbreak{}id}}\\[5pt]

Workflow & P & R & F1 & First returned SQL and candidate count\\

\midrule

No extra model & 25.00 & 100.00 & 40.00 & {\ttfamily\raggedright SELECT District.\allowbreak{}name,\allowbreak{} School\_\allowbreak{}Grades.\allowbreak{}description FROM District JOIN District\_\allowbreak{}School\_\allowbreak{}Grades ON District.\allowbreak{}id = District\_\allowbreak{}School\_\allowbreak{}Grades.\allowbreak{}district\_\allowbreak{}id JOIN School\_\allowbreak{}Grades ON District\_\allowbreak{}School\_\allowbreak{}Grades.\allowbreak{}school\_\allowbreak{}grade\_\allowbreak{}id = School\_\allowbreak{}Grades.\allowbreak{}id}, 7 candidates\\[4pt]

Kimi K3 & 33.33 & 100.00 & 50.00 & Same first query as No extra model, 4 candidates\\[4pt]

Kimi K2.6 & 33.33 & 100.00 & 50.00 & Same first query as No extra model, 4 candidates\\[4pt]

\midrule

\multicolumn{5}{l}{\textit{Recorded intermediate representations}}\\[3pt]

\multicolumn{5}{@{}p{0.98\textwidth}@{}}{\textbf{No extra model.} Atoms a1 = DISTRICT . ID = DISTRICT\_SCHOOL\_GRADES . DISTRICT\_ID. Boolean form a1}\\[4pt]

\multicolumn{5}{@{}p{0.98\textwidth}@{}}{\textbf{Kimi K3.} Atoms a1 = DISTRICT . ID = DISTRICT\_SCHOOL\_GRADES . DISTRICT\_ID. Boolean form a1}\\[4pt]

\multicolumn{5}{@{}p{0.98\textwidth}@{}}{\textbf{Kimi K2.6.} Atoms a1 = DISTRICT . ID = DISTRICT\_SCHOOL\_GRADES . DISTRICT\_ID. Boolean form a1}\\[4pt]

\bottomrule

\end{tabular}

\end{table*}

\begin{table*}[p]

\centering\footnotesize

\setlength{\tabcolsep}{3pt}

\caption{Confirmation feedback. Clear input. Precision, recall, and F1 are percentages computed from the complete returned set against the benchmark references.}

\label{tab:sample-evidence-23}
\begin{tabular}{@{}p{0.19\textwidth}rrrp{0.57\textwidth}@{}}

\toprule

\multicolumn{5}{@{}p{0.98\textwidth}@{}}{\textbf{AMBROSIA 3480}. What are the necessary documents required for this loan.}\\[4pt]

\multicolumn{5}{@{}p{0.98\textwidth}@{}}{\textit{Reference SQL.} {\ttfamily\raggedright Select Loans.\allowbreak{}LoanID,\allowbreak{} EligibilityAndDocuments.\allowbreak{}Documentation\_\allowbreak{}Requirements From EligibilityAndDocuments JOIN Loans ON Loans.\allowbreak{}LoanID = EligibilityAndDocuments.\allowbreak{}LoanId}}\\[5pt]

Workflow & P & R & F1 & First returned SQL and candidate count\\

\midrule

No extra model & 0.00 & 0.00 & 0.00 & {\ttfamily\raggedright SELECT Documentation\_\allowbreak{}Requirements FROM EligibilityAndDocuments}, 3 candidates\\[4pt]

Kimi K3 & 0.00 & 0.00 & 0.00 & Same first query as No extra model, 1 candidate\\[4pt]

Kimi K2.6 & 0.00 & 0.00 & 0.00 & {\ttfamily\raggedright SELECT Documentation\_\allowbreak{}Requirements FROM EligibilityAndDocuments WHERE LoanID = (SELECT MAX(LoanID) FROM Loans)}, 2 candidates\\[4pt]

\midrule

\multicolumn{5}{l}{\textit{Recorded intermediate representations}}\\[3pt]

\multicolumn{5}{@{}p{0.98\textwidth}@{}}{\textbf{No extra model.} Atoms a1 = LOANID = ( SELECT MAX ( LOANID ) FROM LOANS ). Boolean form a1}\\[4pt]

\multicolumn{5}{@{}p{0.98\textwidth}@{}}{\textbf{Kimi K3.} Atoms a1 = LOANID = ( SELECT MAX ( LOANID ) FROM LOANS ). Boolean form a1}\\[4pt]

\multicolumn{5}{@{}p{0.98\textwidth}@{}}{\textbf{Kimi K2.6.} Atoms a1 = LOANID = ( SELECT MAX ( LOANID ) FROM LOANS ). Boolean form a1}\\[4pt]

\bottomrule

\end{tabular}

\end{table*}

\begin{table*}[p]

\centering\footnotesize

\setlength{\tabcolsep}{3pt}

\caption{Confirmation feedback. Clear input. Precision, recall, and F1 are percentages computed from the complete returned set against the benchmark references.}

\label{tab:sample-evidence-24}
\begin{tabular}{@{}p{0.19\textwidth}rrrp{0.57\textwidth}@{}}

\toprule

\multicolumn{5}{@{}p{0.98\textwidth}@{}}{\textbf{AMBROSIA 2950}. Who is the executive chef at the restaurant serving grilled steak with garlic butter.}\\[4pt]

\multicolumn{5}{@{}p{0.98\textwidth}@{}}{\textit{Reference SQL.} {\ttfamily\raggedright SELECT r.\allowbreak{}executive\_\allowbreak{}chef FROM Restaurants r,\allowbreak{} MenuItems m WHERE r.\allowbreak{}restaurant\_\allowbreak{}id = m.\allowbreak{}restaurant\_\allowbreak{}id AND m.\allowbreak{}item\_\allowbreak{}name = 'Steak'}}\\[5pt]

Workflow & P & R & F1 & First returned SQL and candidate count\\

\midrule

No extra model & 100.00 & 100.00 & 100.00 & {\ttfamily\raggedright SELECT DISTINCT Restaurants.\allowbreak{}executive\_\allowbreak{}chef FROM Restaurants JOIN MenuItems ON Restaurants.\allowbreak{}restaurant\_\allowbreak{}id = MenuItems.\allowbreak{}restaurant\_\allowbreak{}id WHERE MenuItems.\allowbreak{}description = 'Grilled steak with garlic butter'}, 4 candidates\\[4pt]

Kimi K3 & 100.00 & 100.00 & 100.00 & Same first query as No extra model, 5 candidates\\[4pt]

Kimi K2.6 & 100.00 & 100.00 & 100.00 & Same first query as No extra model, 4 candidates\\[4pt]

\midrule

\multicolumn{5}{l}{\textit{Recorded intermediate representations}}\\[3pt]

\multicolumn{5}{@{}p{0.98\textwidth}@{}}{\textbf{No extra model.} Atoms a1 = RESTAURANTS . RESTAURANT\_ID = MENUITEMS . RESTAURANT\_ID. Boolean form a1}\\[4pt]

\multicolumn{5}{@{}p{0.98\textwidth}@{}}{\textbf{Kimi K3.} Atoms a1 = RESTAURANTS . RESTAURANT\_ID = MENUITEMS . RESTAURANT\_ID. Boolean form a1}\\[4pt]

\multicolumn{5}{@{}p{0.98\textwidth}@{}}{\textbf{Kimi K2.6.} Atoms a1 = RESTAURANTS . RESTAURANT\_ID = MENUITEMS . RESTAURANT\_ID. Boolean form a1}\\[4pt]

\bottomrule

\end{tabular}

\end{table*}

\begin{table*}[p]

\centering\footnotesize

\setlength{\tabcolsep}{3pt}

\caption{Confirmation feedback. Clear input. Precision, recall, and F1 are percentages computed from the complete returned set against the benchmark references.}

\label{tab:sample-evidence-25}
\begin{tabular}{@{}p{0.19\textwidth}rrrp{0.57\textwidth}@{}}

\toprule

\multicolumn{5}{@{}p{0.98\textwidth}@{}}{\textbf{AMBROSIA 817}. Can you show female research scientists and female laboratory technicians. Show them in one table.}\\[4pt]

\multicolumn{5}{@{}p{0.98\textwidth}@{}}{\textit{Reference SQL.} {\ttfamily\raggedright SELECT Research\_\allowbreak{}Scientists.\allowbreak{}name FROM Research\_\allowbreak{}Scientists WHERE Research\_\allowbreak{}Scientists.\allowbreak{}Gender = "Female" UNION SELECT Laboratory\_\allowbreak{}Technicians.\allowbreak{}name FROM Laboratory\_\allowbreak{}Technicians WHERE Laboratory\_\allowbreak{}Technicians.\allowbreak{}Gender = "Female"}}\\[5pt]

Workflow & P & R & F1 & First returned SQL and candidate count\\

\midrule

No extra model & 33.33 & 100.00 & 50.00 & {\ttfamily\raggedright SELECT name FROM Research\_\allowbreak{}Scientists WHERE Gender = 'Female' UNION ALL SELECT name FROM Laboratory\_\allowbreak{}Technicians WHERE Gender = 'Female'}, 6 candidates\\[4pt]

Kimi K3 & 33.33 & 100.00 & 50.00 & Same first query as No extra model, 6 candidates\\[4pt]

Kimi K2.6 & 25.00 & 100.00 & 40.00 & {\ttfamily\raggedright SELECT * FROM Research\_\allowbreak{}Scientists WHERE Gender = 'Female' UNION ALL SELECT * FROM Laboratory\_\allowbreak{}Technicians WHERE Gender = 'Female'}, 5 candidates\\[4pt]

\midrule

\multicolumn{5}{l}{\textit{Recorded intermediate representations}}\\[3pt]

\multicolumn{5}{@{}p{0.98\textwidth}@{}}{\textbf{No extra model.} Atoms a1 = GENDER = 'Female'. Boolean form a1}\\[4pt]

\multicolumn{5}{@{}p{0.98\textwidth}@{}}{\textbf{Kimi K3.} Atoms a1 = GENDER = 'Female'. Boolean form a1}\\[4pt]

\multicolumn{5}{@{}p{0.98\textwidth}@{}}{\textbf{Kimi K2.6.} Atoms a1 = GENDER = 'Female'. Boolean form a1}\\[4pt]

\bottomrule

\end{tabular}

\end{table*}

\begin{table*}[p]

\centering\footnotesize

\setlength{\tabcolsep}{3pt}

\caption{Confirmation feedback. Ambiguous input. Precision, recall, and F1 are percentages computed from the complete returned set against the benchmark references.}

\label{tab:sample-evidence-26}
\begin{tabular}{@{}p{0.19\textwidth}rrrp{0.57\textwidth}@{}}

\toprule

\multicolumn{5}{@{}p{0.98\textwidth}@{}}{\textbf{AMBIQT 001720}. Count the number of documents that use the PPT template type.}\\[4pt]

\multicolumn{5}{@{}p{0.98\textwidth}@{}}{\textit{Reference SQL 1.} {\ttfamily\raggedright select count(*) from files as t1 join templates as t2 on t1.\allowbreak{}template\_\allowbreak{}id = t2.\allowbreak{}template\_\allowbreak{}id where t2.\allowbreak{}template\_\allowbreak{}type\_\allowbreak{}code = 'PPT'}\par\smallskip \textit{Reference SQL 2.} {\ttfamily\raggedright select count(*) from records as t1 join templates as t2 on t1.\allowbreak{}template\_\allowbreak{}id = t2.\allowbreak{}template\_\allowbreak{}id where t2.\allowbreak{}template\_\allowbreak{}type\_\allowbreak{}code = 'PPT'}}\\[5pt]

Workflow & P & R & F1 & First returned SQL and candidate count\\

\midrule

No extra model & 28.57 & 100.00 & 44.44 & {\ttfamily\raggedright SELECT COUNT(DISTINCT t1.\allowbreak{}document\_\allowbreak{}id) FROM Records AS t1 JOIN templates AS t2 ON t1.\allowbreak{}template\_\allowbreak{}id = t2.\allowbreak{}template\_\allowbreak{}id WHERE t2.\allowbreak{}template\_\allowbreak{}type\_\allowbreak{}code = 'PPT'}, 8 candidates\\[4pt]

Kimi K3 & 33.33 & 100.00 & 50.00 & Same first query as No extra model, 6 candidates\\[4pt]

Kimi K2.6 & 40.00 & 100.00 & 57.14 & Same first query as No extra model, 5 candidates\\[4pt]

\midrule

\multicolumn{5}{l}{\textit{Recorded intermediate representations}}\\[3pt]

\multicolumn{5}{@{}p{0.98\textwidth}@{}}{\textbf{No extra model.} Atoms a1 = T1 . TEMPLATE\_ID = T2 . TEMPLATE\_ID. Boolean form a1}\\[4pt]

\multicolumn{5}{@{}p{0.98\textwidth}@{}}{\textbf{Kimi K3.} Atoms a1 = T1 . TEMPLATE\_ID = T2 . TEMPLATE\_ID. Boolean form a1}\\[4pt]

\multicolumn{5}{@{}p{0.98\textwidth}@{}}{\textbf{Kimi K2.6.} Atoms a1 = T1 . TEMPLATE\_ID = T2 . TEMPLATE\_ID. Boolean form a1}\\[4pt]

\bottomrule

\end{tabular}

\end{table*}

\begin{table*}[p]

\centering\footnotesize

\setlength{\tabcolsep}{3pt}

\caption{Confirmation feedback. Ambiguous input. Precision, recall, and F1 are percentages computed from the complete returned set against the benchmark references.}

\label{tab:sample-evidence-27}
\begin{tabular}{@{}p{0.19\textwidth}rrrp{0.57\textwidth}@{}}

\toprule

\multicolumn{5}{@{}p{0.98\textwidth}@{}}{\textbf{AMBIQT 000309}. List document IDs, document names, and document descriptions for all documents.}\\[4pt]

\multicolumn{5}{@{}p{0.98\textwidth}@{}}{\textit{Reference SQL 1.} {\ttfamily\raggedright select document\_\allowbreak{}id,\allowbreak{} file\_\allowbreak{}name,\allowbreak{} document\_\allowbreak{}description from documents}\par\smallskip \textit{Reference SQL 2.} {\ttfamily\raggedright select document\_\allowbreak{}id,\allowbreak{} record\_\allowbreak{}name,\allowbreak{} document\_\allowbreak{}description from documents}}\\[5pt]

Workflow & P & R & F1 & First returned SQL and candidate count\\

\midrule

No extra model & 66.67 & 100.00 & 80.00 & {\ttfamily\raggedright SELECT document\_\allowbreak{}id,\allowbreak{} file\_\allowbreak{}name,\allowbreak{} document\_\allowbreak{}description FROM documents}, 5 candidates\\[4pt]

Kimi K3 & 66.67 & 100.00 & 80.00 & Same first query as No extra model, 5 candidates\\[4pt]

Kimi K2.6 & 66.67 & 100.00 & 80.00 & Same first query as No extra model, 5 candidates\\[4pt]

\midrule

\multicolumn{5}{l}{\textit{Recorded intermediate representations}}\\[3pt]

\multicolumn{5}{@{}p{0.98\textwidth}@{}}{\textbf{No extra model.} Initial SQL SELECT document\_id, file\_name, document\_description FROM documents,}\\[4pt]

\multicolumn{5}{@{}p{0.98\textwidth}@{}}{\textbf{Kimi K3.} Initial SQL SELECT document\_id, file\_name, document\_description FROM documents,}\\[4pt]

\multicolumn{5}{@{}p{0.98\textwidth}@{}}{\textbf{Kimi K2.6.} Initial SQL SELECT document\_id, file\_name, document\_description FROM documents,}\\[4pt]

\bottomrule

\end{tabular}

\end{table*}

\begin{table*}[p]

\centering\footnotesize

\setlength{\tabcolsep}{3pt}

\caption{Confirmation feedback. Ambiguous input. Precision, recall, and F1 are percentages computed from the complete returned set against the benchmark references.}

\label{tab:sample-evidence-28}
\begin{tabular}{@{}p{0.19\textwidth}rrrp{0.57\textwidth}@{}}

\toprule

\multicolumn{5}{@{}p{0.98\textwidth}@{}}{\textbf{AMBIQT 001542}. What are the ids and makers of all car makers that produce at least 2 models and make more than 3 cars.}\\[4pt]

\multicolumn{5}{@{}p{0.98\textwidth}@{}}{\textit{Reference SQL 1.} {\ttfamily\raggedright select t1.\allowbreak{}id,\allowbreak{} t1.\allowbreak{}maker from car\_\allowbreak{}makers as t1 join model\_\allowbreak{}list as t2 on t1.\allowbreak{}id = t2.\allowbreak{}maker group by t1.\allowbreak{}id having count(*) >= 2 intersect select t1.\allowbreak{}id,\allowbreak{} t1.\allowbreak{}maker from car\_\allowbreak{}makers as t1 join model\_\allowbreak{}list as t2 on t1.\allowbreak{}id = t2.\allowbreak{}maker join car\_\allowbreak{}models as t3 on t2.\allowbreak{}model = t3.\allowbreak{}model group by t1.\allowbreak{}id having count(*) > 3}\par\smallskip \textit{Reference SQL 2.} {\ttfamily\raggedright select t1.\allowbreak{}id,\allowbreak{} t1.\allowbreak{}maker from car\_\allowbreak{}makers as t1 join model\_\allowbreak{}list as t2 on t1.\allowbreak{}id = t2.\allowbreak{}maker group by t1.\allowbreak{}id having count(*) >= 2 intersect select t1.\allowbreak{}id,\allowbreak{} t1.\allowbreak{}maker from car\_\allowbreak{}makers as t1 join model\_\allowbreak{}list as t2 on t1.\allowbreak{}id = t2.\allowbreak{}maker join vehicle\_\allowbreak{}names as t3 on t2.\allowbreak{}model = t3.\allowbreak{}model group by t1.\allowbreak{}id having count(*) > 3}}\\[5pt]

Workflow & P & R & F1 & First returned SQL and candidate count\\

\midrule

No extra model & 0.00 & 0.00 & 0.00 & {\ttfamily\raggedright SELECT t1.\allowbreak{}id,\allowbreak{} t1.\allowbreak{}maker FROM car\_\allowbreak{}makers AS t1 JOIN model\_\allowbreak{}list AS t2 ON t1.\allowbreak{}id = t2.\allowbreak{}maker JOIN car\_\allowbreak{}models AS t3 ON t2.\allowbreak{}model = t3.\allowbreak{}model JOIN cars\_\allowbreak{}data AS t4 ON t3.\allowbreak{}makeid = t4.\allowbreak{}id GROUP BY t1.\allowbreak{}id,\allowbreak{} t1.\allowbreak{}maker HAVING COUNT(DISTINCT t2.\allowbreak{}modelid) >= 2 AND COUNT(DISTINCT t4.\allowbreak{}id) > 3}, 5 candidates\\[4pt]

Kimi K3 & 0.00 & 0.00 & 0.00 & {\ttfamily\raggedright SELECT t1.\allowbreak{}id,\allowbreak{} t1.\allowbreak{}maker FROM car\_\allowbreak{}makers AS t1 JOIN model\_\allowbreak{}list AS t2 ON t1.\allowbreak{}id = t2.\allowbreak{}maker JOIN vehicle\_\allowbreak{}names AS t3 ON t2.\allowbreak{}model = t3.\allowbreak{}model JOIN cars\_\allowbreak{}data AS t4 ON t3.\allowbreak{}makeid = t4.\allowbreak{}id GROUP BY t1.\allowbreak{}id,\allowbreak{} t1.\allowbreak{}maker HAVING COUNT(DISTINCT t2.\allowbreak{}model) >= 2 AND COUNT(DISTINCT t4.\allowbreak{}id) > 3}, 4 candidates\\[4pt]

Kimi K2.6 & 0.00 & 0.00 & 0.00 & {\ttfamily\raggedright SELECT t1.\allowbreak{}id,\allowbreak{} t1.\allowbreak{}maker FROM car\_\allowbreak{}makers AS t1 JOIN model\_\allowbreak{}list AS t2 ON t1.\allowbreak{}id = t2.\allowbreak{}maker JOIN vehicle\_\allowbreak{}names AS t3 ON t1.\allowbreak{}id = t3.\allowbreak{}makeid GROUP BY t1.\allowbreak{}id,\allowbreak{} t1.\allowbreak{}maker HAVING COUNT(DISTINCT t2.\allowbreak{}model) >= 2 AND COUNT(DISTINCT t3.\allowbreak{}model) > 3}, 3 candidates\\[4pt]

\midrule

\multicolumn{5}{l}{\textit{Recorded intermediate representations}}\\[3pt]

\multicolumn{5}{@{}p{0.98\textwidth}@{}}{\textbf{No extra model.} Atoms a1 = T1 . ID = T2 . MAKER. Boolean form a1}\\[4pt]

\multicolumn{5}{@{}p{0.98\textwidth}@{}}{\textbf{Kimi K3.} Atoms a1 = T1 . ID = T2 . MAKER. Boolean form a1}\\[4pt]

\multicolumn{5}{@{}p{0.98\textwidth}@{}}{\textbf{Kimi K2.6.} Atoms a1 = T1 . ID = T2 . MAKER. Boolean form a1}\\[4pt]

\bottomrule

\end{tabular}

\end{table*}

\begin{table*}[p]

\centering\footnotesize

\setlength{\tabcolsep}{3pt}

\caption{Confirmation feedback. Ambiguous input. Precision, recall, and F1 are percentages computed from the complete returned set against the benchmark references.}

\label{tab:sample-evidence-29}
\begin{tabular}{@{}p{0.19\textwidth}rrrp{0.57\textwidth}@{}}

\toprule

\multicolumn{5}{@{}p{0.98\textwidth}@{}}{\textbf{AMBIQT 002504}. Find the id, last name and cell phone of the professionals who live in the state of Indiana or have performed more than two treatments.}\\[4pt]

\multicolumn{5}{@{}p{0.98\textwidth}@{}}{\textit{Reference SQL 1.} {\ttfamily\raggedright select professional\_\allowbreak{}id,\allowbreak{} last\_\allowbreak{}name,\allowbreak{} cell\_\allowbreak{}number from handlers where state = 'Indiana' union select t1.\allowbreak{}professional\_\allowbreak{}id,\allowbreak{} t1.\allowbreak{}last\_\allowbreak{}name,\allowbreak{} t1.\allowbreak{}cell\_\allowbreak{}number from handlers as t1 join treatments as t2 on t1.\allowbreak{}professional\_\allowbreak{}id = t2.\allowbreak{}professional\_\allowbreak{}id group by t1.\allowbreak{}professional\_\allowbreak{}id having count(*) > 2}\par\smallskip \textit{Reference SQL 2.} {\ttfamily\raggedright select professional\_\allowbreak{}id,\allowbreak{} last\_\allowbreak{}name,\allowbreak{} cell\_\allowbreak{}number from caregivers where state = 'Indiana' union select t1.\allowbreak{}professional\_\allowbreak{}id,\allowbreak{} t1.\allowbreak{}last\_\allowbreak{}name,\allowbreak{} t1.\allowbreak{}cell\_\allowbreak{}number from caregivers as t1 join treatments as t2 on t1.\allowbreak{}professional\_\allowbreak{}id = t2.\allowbreak{}professional\_\allowbreak{}id group by t1.\allowbreak{}professional\_\allowbreak{}id having count(*) > 2}}\\[5pt]

Workflow & P & R & F1 & First returned SQL and candidate count\\

\midrule

No extra model & 40.00 & 100.00 & 57.14 & {\ttfamily\raggedright SELECT professional\_\allowbreak{}id,\allowbreak{} last\_\allowbreak{}name,\allowbreak{} cell\_\allowbreak{}number FROM Caregivers WHERE state = 'Indiana' OR professional\_\allowbreak{}id IN (SELECT professional\_\allowbreak{}id FROM treatments GROUP BY professional\_\allowbreak{}id HAVING COUNT(*) > 2)}, 5 candidates\\[4pt]

Kimi K3 & 50.00 & 100.00 & 66.67 & Same first query as No extra model, 4 candidates\\[4pt]

Kimi K2.6 & 0.00 & 0.00 & 0.00 & {\ttfamily\raggedright SELECT t1.\allowbreak{}professional\_\allowbreak{}id,\allowbreak{} t1.\allowbreak{}last\_\allowbreak{}name,\allowbreak{} t1.\allowbreak{}cell\_\allowbreak{}number FROM (SELECT professional\_\allowbreak{}id,\allowbreak{} last\_\allowbreak{}name,\allowbreak{} cell\_\allowbreak{}number,\allowbreak{} state FROM Caregivers UNION SELECT professional\_\allowbreak{}id,\allowbreak{} last\_\allowbreak{}name,\allowbreak{} cell\_\allowbreak{}number,\allowbreak{} state FROM Handlers) AS t1 WHERE t1.\allowbreak{}state = 'Indiana' OR t1.\allowbreak{}professional\_\allowbreak{}id IN (SELECT t2.\allowbreak{}professional\_\allowbreak{}id FROM treatments AS t2 GROUP BY t2.\allowbreak{}professional\_\allowbreak{}id HAVING COUNT(*) > 2)}, 4 candidates\\[4pt]

\midrule

\multicolumn{5}{l}{\textit{Recorded intermediate representations}}\\[3pt]

\multicolumn{5}{@{}p{0.98\textwidth}@{}}{\textbf{No extra model.} Atoms a1 = PROFESSIONAL\_ID IN ( SELECT PROFESSIONAL\_ID FROM TREATMENTS GROUP BY PROFESSIONAL\_ID HAVING COUNT ( * ) > 2 ), a2 = STATE = 'Indiana'. Boolean form (a2 OR a1)}\\[4pt]

\multicolumn{5}{@{}p{0.98\textwidth}@{}}{\textbf{Kimi K3.} Atoms a1 = PROFESSIONAL\_ID IN ( SELECT PROFESSIONAL\_ID FROM TREATMENTS GROUP BY PROFESSIONAL\_ID HAVING COUNT ( * ) > 2 ), a2 = STATE = 'Indiana'. Boolean form (a2 OR a1)}\\[4pt]

\multicolumn{5}{@{}p{0.98\textwidth}@{}}{\textbf{Kimi K2.6.} Atoms a1 = T1 . PROFESSIONAL\_ID IN ( SELECT T2 . PROFESSIONAL\_ID FROM TREATMENTS AS T2 GROUP BY T2 . PROFESSIONAL\_ID HAVING COUNT ( * ) > 2 ), a2 = T1 . STATE = 'Indiana'. Boolean form (a2 OR a1)}\\[4pt]

\bottomrule

\end{tabular}

\end{table*}

\begin{table*}[p]

\centering\footnotesize

\setlength{\tabcolsep}{3pt}

\caption{Confirmation feedback. Ambiguous input. Precision, recall, and F1 are percentages computed from the complete returned set against the benchmark references.}

\label{tab:sample-evidence-30}
\begin{tabular}{@{}p{0.19\textwidth}rrrp{0.57\textwidth}@{}}

\toprule

\multicolumn{5}{@{}p{0.98\textwidth}@{}}{\textbf{AMBIQT 002572}. Which dogs are of the rarest breed. Show their names and treatment dates.}\\[4pt]

\multicolumn{5}{@{}p{0.98\textwidth}@{}}{\textit{Reference SQL 1.} {\ttfamily\raggedright select t1.\allowbreak{}name,\allowbreak{} t2.\allowbreak{}date\_\allowbreak{}of\_\allowbreak{}treatment from canines as t1 join treatments as t2 on t1.\allowbreak{}dog\_\allowbreak{}id = t2.\allowbreak{}dog\_\allowbreak{}id where t1.\allowbreak{}breed\_\allowbreak{}code = (select breed\_\allowbreak{}code from canines group by breed\_\allowbreak{}code order by count(*) asc limit 1)}\par\smallskip \textit{Reference SQL 2.} {\ttfamily\raggedright select t1.\allowbreak{}name,\allowbreak{} t2.\allowbreak{}date\_\allowbreak{}of\_\allowbreak{}treatment from pooches as t1 join treatments as t2 on t1.\allowbreak{}dog\_\allowbreak{}id = t2.\allowbreak{}dog\_\allowbreak{}id where t1.\allowbreak{}breed\_\allowbreak{}code = (select breed\_\allowbreak{}code from pooches group by breed\_\allowbreak{}code order by count(*) asc limit 1)}}\\[5pt]

Workflow & P & R & F1 & First returned SQL and candidate count\\

\midrule

No extra model & 0.00 & 0.00 & 0.00 & {\ttfamily\raggedright SELECT t1.\allowbreak{}name,\allowbreak{} t2.\allowbreak{}date\_\allowbreak{}of\_\allowbreak{}treatment FROM Pooches AS t1 JOIN Treatments AS t2 ON t1.\allowbreak{}dog\_\allowbreak{}id = t2.\allowbreak{}dog\_\allowbreak{}id WHERE t1.\allowbreak{}breed\_\allowbreak{}code IN (SELECT breed\_\allowbreak{}code FROM Pooches GROUP BY breed\_\allowbreak{}code HAVING COUNT(*) = (SELECT MIN(cnt) FROM (SELECT COUNT(*) AS cnt FROM Pooches GROUP BY breed\_\allowbreak{}code)))}, 5 candidates\\[4pt]

Kimi K3 & 0.00 & 0.00 & 0.00 & Same first query as No extra model, 5 candidates\\[4pt]

Kimi K2.6 & 0.00 & 0.00 & 0.00 & Same first query as No extra model, 5 candidates\\[4pt]

\midrule

\multicolumn{5}{l}{\textit{Recorded intermediate representations}}\\[3pt]

\multicolumn{5}{@{}p{0.98\textwidth}@{}}{\textbf{No extra model.} Atoms a1 = T1 . DOG\_ID = T2 . DOG\_ID. Boolean form a1}\\[4pt]

\multicolumn{5}{@{}p{0.98\textwidth}@{}}{\textbf{Kimi K3.} Atoms a1 = T1 . DOG\_ID = T2 . DOG\_ID. Boolean form a1}\\[4pt]

\multicolumn{5}{@{}p{0.98\textwidth}@{}}{\textbf{Kimi K2.6.} Atoms a1 = T1 . DOG\_ID = T2 . DOG\_ID. Boolean form a1}\\[4pt]

\bottomrule

\end{tabular}

\end{table*}

\end{document}